\documentclass[oneside,11pt]{article}%
\usepackage{amsfonts}
\usepackage{amssymb}
\usepackage{amsmath}
\usepackage{amsthm}
\usepackage{float}
\usepackage{graphicx}
\usepackage[dvipsnames]{xcolor}
\usepackage{url}
\usepackage[colorlinks=true,linkcolor=RawSienna,citecolor=RawSienna,urlcolor=RawSienna,pdfstartview=FitB]%
{hyperref}
\usepackage[left=1.1in,right=1.1in,top=1.1in,bottom=1.1in]{geometry}
\usepackage{natbib}
\usepackage{longtable,lscape}
\usepackage[onehalfspacing]{setspace}
\usepackage{hyphenat}%

\newtheorem{theorem}{Theorem}[section]
\newtheorem{corollary}{Corollary}[section]

\newtheorem{lemma}{Lemma}[section]

\allowdisplaybreaks

\numberwithin{equation}{section}
\begin{document}

\title{\vspace{-1.2cm}{Consistent estimation of the proportion of false nulls and FDR
for adaptive multiple testing Normal means under weak dependence}}
\author{Xiongzhi Chen\thanks{Corresponding author. Department of Mathematics and
Statistics, Washington State University, Pullman, WA 99164, USA; E-mail:
\texttt{xiongzhi.chen@wsu.edu}} }
\date{}
\maketitle

\begin{abstract}
We consider multiple testing means of many dependent Normal random variables
that do not necessarily follow a joint Normal distribution. Under weak
dependence, we show the uniform consistency of proportion estimators that are
constructed as solutions to Lebesgue-Stieltjes equations for the setting of a
point, bounded and one-sided null, respectively, and characterize via the
index of weak dependence the sparsest proportion these estimators can
consistently estimate. On the other hand, under a principal correlation
structure and employing a suitable definition of p-value for composite null
hypotheses, we show that three key empirical processes induced by a
single-step multiple testing procedure (MTP) satisfy the strong law of large
numbers for testing each of the three types of nulls. Further, under this
structure and for testing a point null and a one-sided null respectively, we
construct an adaptive single-step MTP that employs a proportion estimator
mentioned earlier, and show that the false discovery proportion of this
procedure satisfies the weak law of large numbers and hence consistently
estimates the false discovery rate of the procedure. In addition, we report
some findings on the estimators of Jin and of Meinshausen and Rice of the
proportion of false nulls in the critically and very sparse
regimes under weak dependence and model misspecifications, respectively.
\end{abstract}
\tableofcontents

\section{Introduction}

\label{sec1intro}

We consider the multiple testing scenario where the means of many dependent
Normal random variables are simultaneously tested. This scenario is a classic
subject of theoretical study in statistics. It has also been conducted when
implementing a strategy called \textquotedblleft marginal regression followed
by multiple testing\textquotedblright, where many marginal regression models
are obtained, a null hypothesis is that a regression coefficient is zero, the
associated test statistic follows a Normal distribution and has mean zero
under the null hypothesis, and all regression coefficients are simultaneous
assessed; see, e.g., \cite{Owen:2005,Fan:2012,Schwartzman:2015}.

Due to the large number of null hypotheses to test simultaneously, it is often
desirable to control the false discovery rate (FDR) in order to identify a set
of regression coefficients for further study. However, the Normally
distributed test statistics can be highly dependent, and the behavior of the
false discovery proportion (FDP) of a multiple testing procedure (MTP) such as
the Benjamini-Hochberg (BH) procedure of \cite{Benjamini:1995} is unstable and
sometimes even unpredictable; see, e.g., \cite{Finner:2007}. Considerable
efforts have been carried out to identify dependence structures under which a
BH-type step-up procedure is conservative, either non-asymptotically or
asymptotically, as by, e.g.,
\cite{Benjamini:2001,Storey:2004,Finner:2007,farcomeni:2007}. Parallel to
these efforts, another line of research studies a single-step MTP and aims at
asymptotic FDR control under dependence; see, e.g.,
\cite{Friguet:2009,Schwartzman:2011,Fan:2012,Delattre:2016,Fan:2017,Chen:2014SLLN}%
. Finally, adaptive MTPs that incorporate a suitably accurate estimate of the
proportion of true nulls can be more powerful than their non-adaptive
counterparts when this proportion is not very close one; see, e.g.,
\cite{Storey:2004,Benjamini:2006,Blanchard:2009,Nandi:2018}.

However, the above mentioned works are for multiple testing a point null
hypothesis and are not for adaptive single-step MTPs that incorporate
proportion estimators. Even though Bayesian mixture models have been used,
e.g., by \cite{Storey:2003c,Genovese:2004,Chi:2010}, to route testing
composite nulls back to testing point nulls, we will take a frequentist
perspective here since Bayesian mixture models often unrealistically force
test statistics to be marginally identically distributed. So, there is a need
to fill the gap of consistently estimating the proportion of false nulls
(i.e., one minus the proportion of true nulls) and consistently estimating the
FDR of an adaptive single-step MTP under dependence and for multiple testing
composite nulls. This motivates our work.

\subsection{Summary of main contributions}

In this article, we study the FDP of an adaptive single-step MTP when each
pair of test statistics follows a bivariate Normal distribution with known
variances. Firstly, we consider a single-step MTP and show that, for the
setting of a point null, bounded null and one-sided null respectively, the
empirical processes of the number of false rejections, the total number of
rejections and the FDP all satisfy the strong law of large numbers when the
Normal random variables have a principal correlation structure (PCS).
Secondly, we show that, for testing a point null and a one-sided null
respectively, the FDP of the adaptive single-step MTP that employs a
consistent proportion estimator of the proportion of false nulls satisfies the
weak law of large numbers. So, the FDR of the procedure can be consistently
estimated by its FDP, and the rejection threshold of the procedure can be
explicitly determined to asymptotically control its FDR at a desired level.
Thirdly, for the setting of a point null, bounded null and one-sided null
respectively, we show that the proportion estimators of \cite{Chen:2018a} and
\cite{Chen:2019a} are uniformly consistent when the Normal random variables
are weakly dependent. This includes an extension in the setting of a point
null Jin's estimator of \cite{Jin:2008} to the case where Normal random
variables have heterogeneous variances and are weakly dependent.

Further, our simulation study investigates the performances of Jin's estimator
and the \textquotedblleft MR\textquotedblright\ estimator of
\cite{Meinshausen:2006} for three types of dependence in the setting of a
point null, none of which has been considered in the simulation studies of
\cite{Meinshausen:2006}, \cite{Jin:2007} or \cite{Jin:2008}. In particular, we
have observed that Jin's estimator is overall more accurate than the MR
estimator, that Jin's estimator can have reasonable performance in the
critically and very sparse regimes even though it has not been theoretically
proven to be consistent in these regimes, and that both estimators are
relatively robust to violations of assumptions needed for their consistency.

\subsection{Notations and conventions}

The notations and conventions we will use throughout are stated as follows:
$\mathbb{P}$ is the probability measure, and $\mathbb{E}$, $\mathbb{V}$ and
$\mathsf{cov}$ are the expectation, variance and covariance operators with
respect to $\mathbb{P}$; $\mathbf{w}\sim\mathcal{N}_{m}\left(  \mathbf{u}%
,\mathbf{S}\right)  $ denotes that $\mathbf{w}$ is an $m$-dimensional Normally
distributed random vector with mean vector $\mathbf{u}$ and covariance matrix
$\mathbf{S}$, and $\Phi$ the cumulative distribution function (CDF) of
$X\sim\mathcal{N}_{1}\left(  0,1\right)  $; for two sequences $\left\{
a_{m}\right\}  _{m\geq1}$ and $\left\{  b_{m}\right\}  _{m\geq1}$, the
\textquotedblleft big O\textquotedblright\ notation $a_{m}=O\left(
b_{m}\right)  $ means that $\left\vert a_{m}\right\vert \leq M\left\vert
b_{m}\right\vert $ for some constant $M>0$ for all sufficiently large $m$; for
a matrix $\mathbf{A}$, let $\left\Vert \mathbf{A}\right\Vert _{1}=\sum
_{i,j}\left\vert \mathbf{A}\left(  i,j\right)  \right\vert$ be its $l_1$-norm; $1_{A}$ is the
indicator of a set $A$; $\left[  x\right]  $ denotes the integer part of a
real number $x$; $\Re$ denotes the real part of a complex number, and
$\iota=\sqrt{-1}$.

\subsection{Organization of article}

The rest of the article is organized as follows. In \autoref{secModel} we
state the model for multiple testing Normal means, the definition of PCS and
its implications on the stability of an MTP, and the methodology of adaptive
single-step MTP and consistent estimation of its FDR. In
\autoref{secAdaptiveTesting} we introduce the proportion estimators of
\cite{Chen:2018a} and \cite{Chen:2019a} in the setting of Normal random
variables and show their uniform consistency under weak dependence. We provide
in \autoref{SecNumericalStudies} a simulation study on these proportion
estimators for multiple testing a point, bounded and one-sided null
respectively and on an adaptive single-step MTP that employs such an estimator
for multiple testing a point null. We end the article with a discussion in
\autoref{SecConcAndDisc} and relegate all proofs into the appendix.

\section{Adaptive multiple testing Normal means}

\label{secModel}

The main purpose of this section is to present the framework for adaptive
multiple testing the means of Normal random variables and some key and new
results on the stability of a single-step MTP under PCS. Specifically, we will
state in \autoref{secMTP} the problem of multiple testing Normal means, in
\autoref{secPCSDef} the relationship between PCS and the stability of three
key empirical processes induced by an MTP, and in \autoref{ConstFDREst}
adaptive single-step multiple testing and consistent estimation of FDR under PCS.

\subsection{Multiple testing, proportion of false nulls and p-value}

\label{secMTP}

Let $\Theta_{0}$ be a subset of the set of real numbers $\mathbb{R}$ that has
a non-empty interior and non-empty complement $\Theta_{1}=\mathbb{R}%
\setminus\Theta_{0}$. For each $i\in\left\{  1,\ldots,m\right\}  $, let
$z_{i}\sim\mathcal{N}_{1}\left(  \mu_{i},\sigma_{ii}\right)  $, such that, for
some integer $m_{0}$ between $0$ and $m$, $\mu_{i}\in\Theta_{0}$ for each
$i\in\left\{  1,\ldots,m_{0}\right\}  $ and $\mu_{i}\in\Theta_{1}$ for each
$i\in\left\{  m_{0}+1,\ldots,m\right\}  $. Consider simultaneously testing the
null hypothesis $H_{i0}:\mu_{i}\in\Theta_{0}$ versus the alternative
hypothesis $H_{i1}:\mu_{i}\in\Theta_{1}$ for all $i\in\left\{  1,\ldots
,m\right\}  $. For testing the means of Normal random variables, without loss
of generality, we set $\Theta_{0}=\left\{  0\right\}  $ as the
\textquotedblleft point null\textquotedblright, $\Theta_{0}=\left(
-\infty,0\right)  $ the \textquotedblleft one-sided null\textquotedblright,
and $\Theta_{0}=\left(  a,b\right)  $ for fixed, finite $a,b$ with $a<b$ a
\textquotedblleft bounded null\textquotedblright. The case where $\Theta
_{0}=\left\{  0\right\}  $ is the classic \textquotedblleft Normal means
problem\textquotedblright\ that has been studied by many; see, e.g.,
\cite{Donoho:2004} and \cite{WassermanBook:2006}. In contrast, a one-sided or
bounded null is often encountered in genomics or genetics when a practitioner
is looking for down- or up-regulation for genes or assessing the range of
fold-changes in gene expressions.

Let $I_{0,m}$ $=\left\{  i\in\left\{  1,\ldots,m\right\}  :\mu_{i}\in
\Theta_{0}\right\}  $ and $I_{1,m}=\left\{  i\in\left\{  1,\ldots,m\right\}
:\mu_{i}\in\Theta_{1}\right\}  $. Then the cardinality of $I_{0,m}$ is $m_{0}%
$, the proportion of true nulls (\textquotedblleft null
proportion\textquotedblright\ for short) is defined as $\pi_{0,m}=m^{-1}m_{0}%
$, and the proportion of false nulls (\textquotedblleft alternative
proportion\textquotedblright\ for short) $\pi_{1,m}=1-\pi_{0,m}$. In other
words,%
\begin{equation}
\pi_{1,m}=m^{-1}\left\vert \left\{  i\in\left\{  1,...,m\right\}  :\mu_{i}%
\in\Theta_{1}\right\}  \right\vert \label{defPiOne}%
\end{equation}
is the proportion of Normal random variables whose means are in $\Theta_{1}$.
The proportions $\pi_{0,m}$ and $\pi_{1,m}$ appear in upper bounds on FDR and
false nondiscovery rate (FNR, \cite{Genovese:2002}), and suitable estimates of
these proportions are used to construct adaptive FDR procedures by, e.g.,
\cite{Storey:2004,Sarkar:2006,Benjamini:2006,Blanchard:2009,Chen:2018d,Nandi:2018}%
.

For $\Theta_{0}=\left\{  0\right\}  $, the one-sided p-value for $z_{i}$ is
defined as \textquotedblleft right-tailed\textquotedblright\ $p_{i}%
=1-F_{0}\left(  z_{i}\right)  $ or \textquotedblleft
left-tailed\textquotedblright\ $p_{i}=F_{0}\left(  z_{i}\right)  $, and the
two-sided p-value for $z_{i}$ as $p_{i}=2F_{0}\left(  -\left\vert
z_{i}\right\vert \right)  $, where $F_{0}$ is the CDF of $X\sim\mathcal{N}%
_{1}\left(  0,\sigma_{ii}\right)  $. In contrast, if $\Theta_{0}$ is not a
singleton, i.e., if $\Theta_{0}$ is a composite null, then a p-value for
$z_{i}$ can be defined as%
\begin{equation}
p_{i}=\sup\nolimits_{\mu\in\Theta_{0}}F_{\mu}\left(  z_{i}\right)  \text{
where }F_{\mu}\text{ is the CDF of }X\sim\mathcal{N}_{1}\left(  \mu
,\sigma_{ii}\right)  ; \label{defPvalComp}%
\end{equation}
see, e.g., Definition 2.1 in Chapter of \cite{Dickhaus}. It is widely known
that definition (\ref{defPvalComp}) can be very conservative in terms of power
and hard to compute numerically, which is perhaps the main obstacle for its
wide usage. For the one-sided null $\Theta_{0}=\left(  -\infty,0\right)  $ and
when $z_{i}$'s have identical variances, the family of densities of $\left\{
F_{\mu_{i}}\right\}  _{i=1}^{m}$ forms a monotone likelihood ratio family and
$p_{i}$ in (\ref{defPvalComp}) can be set as $p_{i}=1-F_{0}\left(
z_{i}\right)  $ using the concept of uniformly most powerful test for testing
a single hypothesis; see, e.g., Chapter 3 of \cite{Lehmann:2005}. We call this
definition a \textquotedblleft classic definition\textquotedblright\ of
p-value for the one-sided null $\Theta_{0}=\left(  -\infty,0\right)  $. Note
that other definitions of p-value for a composite null have been provided by,
e.g., \cite{Bayarri:2000}, \cite{Chi:2010} and \cite{Dickhaus:2013}. However,
studying which among all these definitions is better under what circumstances
is not our focus here.

\subsection{Principal correlation structure and stability of multiple testing}

\label{secPCSDef}

Consider the single-step MTP with a rejection threshold $\tau\in\left(
0,1\right)  $ that rejects $H_{i0}:\mu_{i}\in\Theta_{0}$ if and only if
$p_{i}\leq\tau$. Then $R_{m}\left(  \tau\right)  =\sum_{i=1}^{m}1_{\left\{
p_{i}\leq\tau\right\}  }$ is the number of rejections and $V_{m}\left(
\tau\right)  =\sum_{i\in I_{0,m}}1_{\left\{  p_{i}\leq\tau\right\}  }$ the
number of false discoveries. Further, the FDP and FDR of the MTP are
respectively
\[
\mathrm{FDP}_{m}\left(  \tau\right)  =\frac{V_{m}\left(  \tau\right)  }%
{\max\left\{  R_{m}\left(  \tau\right)  ,1\right\}  }\text{ \ \ \ and
\ \ \ \ }\mathrm{FDR}_{m}\left(  \tau\right)  =\mathbb{E}\left[
\mathrm{FDP}_{m}\left(  \tau\right)  \right]  .
\]
When $m$ is large, we aim to control the FDR of the MTP at a given level
$\alpha\in\left(  0,1\right)  $ by choosing an appropriate $\tau$ or to
estimate the FDR of the MTP for a given $\tau$.

Let $\mathbf{z}=\left(  z_{1},\ldots,z_{m}\right)  ^{\top}$ and let
$\mathbf{\Sigma}=\left(  \sigma_{ij}\right)  $ and $\mathbf{R}=\left(
\rho_{ij}\right)  _{m\times m}$ be the covariance matrix and correlation
matrix of $\mathbf{z}$ respectively. If
\begin{equation}
m^{-2}\left\Vert \mathbf{R}\right\Vert _{1}=O\left(  m^{-\beta}\right)  \text{
\ for some \ }\beta>0, \label{24}%
\end{equation}
we say that $\mathbf{z}$ has a \textquotedblleft principal correlation
structure (PCS)\textquotedblright\ as defined by \cite{Chen:2014SLLN} and call
$\beta$ the PCS index. PCS says that the sum of the absolute correlations
among all components of the random vector $\mathbf{z}$ accumulates at a slower
rate than the squared dimensionality of $\mathbf{z}$. Note that $\beta$ in
(\ref{24}) has to be no larger than $1$. The components of $\mathbf{z}$ for
which $\lim_{m\rightarrow\infty}m^{-2}\left\Vert \mathbf{R}\right\Vert _{1}=0$
are referred to as \textquotedblleft weakly correlated\textquotedblright\ by
\cite{Schwartzman:2015}. In view of this, PCS is a more quantified version of
weak correlation. A related concept is \textquotedblleft weak
dependence\textquotedblright\ defined by \cite{Fan:2012}, for which $\left\{
z_{i}\right\}  _{i=1}^{m}$ are \textquotedblleft weakly
dependent\textquotedblright\ if%
\begin{equation}
m^{-2}\left\Vert \mathbf{\Sigma}\right\Vert _{1}=O\left(  m^{-\delta}\right)
\text{ \ for \ some \ }\delta>0. \label{eq23}%
\end{equation}
We refer to $\delta$ in (\ref{eq23}) as the \textquotedblleft index of weak
dependence\textquotedblright, and $\delta>1$ can happen when some $\sigma
_{ii}$'s are zero. Let%
\[
\sigma_{\left(  1\right)  }=\min_{1\leq i\leq m}\sigma_{ii}\ \ \ \text{and
\ }\sigma_{\left(  m\right)  }=\max_{1\leq i\leq m}\sigma_{ii}.
\]
Since (\ref{eq23}) implies (\ref{24}) when $\lim_{m\rightarrow\infty}%
\sigma_{\left(  1\right)  }>0$ and $\lim_{m\rightarrow\infty}\sigma_{\left(
m\right)  }<\infty$ whereas (\ref{24}) implies (\ref{eq23}) when
$\lim_{m\rightarrow\infty}\sigma_{\left(  m\right)  }<\infty$, neither PCS nor
weak dependence subsumes the other in terms of set inclusion.

It is easy to verify that PCS subsumes strongly mixing and short-range
dependence considered in \cite{Jin:2007}. Specifically, we can construct
$\mathbf{z}$ such that (\ref{24}) holds and that for $k=10^{-1}\sqrt{m}$, all
entries of the covariance matrix of the $2k$ components $z_{j}$ and
$z_{m+1-j}$ for $1\leq j\leq k$ are uniformly bounded in $m$ from below by a
positive constant, and thus the sequence $\left\{  z_{i}\right\}  _{i=1}^{m}$
is neither strongly mixing nor have short-range dependence; see also the
\textquotedblleft Long Range\textquotedblright\ dependence in the simulation
study in \autoref{SecNumericalStudies}. Further, condition (\ref{24})
complements several mixing conditions that were used by, e.g.,
\cite{farcomeni:2007}, as explained below. Let $m>1$. Consider a sequence
$\left\{  z_{i}\right\}  _{i=1}^{m}$, such that $z_{1}$ and $z_{m}$ have
non-zero, constant variances and covariance and that $z_{i}$ and $z_{j}$ with
$i\neq j$ and $i,j\neq1$ or $m$ are independent. Then, $\left\{
z_{i}\right\}  _{i=1}^{m}$ cannot be mixing but satisfies (\ref{24}). In
contrast, a sequence $\left\{  z_{i}\right\}  _{i=1}^{m}$ such that the
correlation between $z_{i}$ and $z_{j},i\neq j$ is $1/\ln{m}$ can be mixing
but cannot satisfy (\ref{24}).

One implication of PCS is the following strong law of large numbers for three
empirical processes induced by a single-step MTP for testing a point null
hypothesis $\Theta_{0}=\left\{  0\right\}  $, shown by \cite{Chen:2014SLLN}:

\begin{theorem}
\label{prop:SLLNMarginal}Consider the point null $\Theta_{0}=\left\{
0\right\}  $ and its associated right-tailed or two-sided p-values $\left\{
p_{i}\right\}  _{i=1}^{m}$ for $\left\{  z_{i}\right\}  _{i=1}^{m}$. If each
pair $\left(  z_{i},z_{j}\right)  $ with $i\neq j$ is bivariate Normal, then
there exists a constant $C>0$ such that%
\begin{equation}
\left\vert \mathrm{cov}\big(1_{\left\{  p_{i}\leq\tau\right\}  },1_{\left\{
p_{j}\leq\tau\right\}  }\big)\right\vert \leq C\left\vert \rho_{ij}\right\vert
\text{ for all }i\neq j\text{.} \label{IneqA}%
\end{equation}
If in addition (\ref{24}) holds, then%
\begin{equation}
\lim_{m\rightarrow\infty}m^{-1}\left\vert R_{m}\left(  \tau\right)
-\mathbb{E}\left[  R_{m}\left(  \tau\right)  \right]  \right\vert
=0=\lim_{m_{0}\rightarrow\infty}m_{0}^{-1}\left\vert V_{m}\left(  \tau\right)
-\mathbb{E}\left[  V_{m}\left(  \tau\right)  \right]  \right\vert \text{
almost surely}. \label{SLLNAB}%
\end{equation}
If further $\liminf_{m\rightarrow\infty}m^{-1}R_{m}\left(  t\right)  >0$,
then
\begin{equation}
\lim_{m\rightarrow\infty}\left\vert \mathrm{FDP}_{m}\left(  \tau\right)
-\mathbb{E}\left[  \mathrm{FDP}_{m}\left(  \tau\right)  \right]  \right\vert
=0\text{ almost surely.} \label{SLLNC}%
\end{equation}

\end{theorem}

Condition (\ref{24}) provides a way to check if the \textquotedblleft weak
dependence\textquotedblright\ assumption, which was proposed by
\cite{Storey:2004} and requires (\ref{SLLNAB}), holds, and the condition
$\liminf_{m\rightarrow\infty}m^{-1}R_{m}\left(  t\right)  >0$ is a very
important sufficient condition, without which a linear step-up MTP may fail to
control its FDR\ asymptotically even if $\lim_{m_{0}\rightarrow\infty}%
m_{0}^{-1}\left\vert V_{m}\left(  \tau\right)  -\mathbb{E}\left[  V_{m}\left(
\tau\right)  \right]  \right\vert =0$ holds almost surely; see
\cite{Gontscharuk:2013}. \autoref{prop:SLLNMarginal} asserts that, for
multiple testing whether the means of Normal random variables are zero or not
under PCS, if there is always a positive proportion of rejected nulls, then
the FDR of the MTP can be estimated arbitrarily well by its FDP as the number
of tests increases. However, it does not necessarily hold under weak
dependence or weak correlation. Note that the original statement of
\autoref{prop:SLLNMarginal} used the condition $\liminf_{m\rightarrow\infty
}\pi_{0,m}>0$, which implies $\liminf_{m\rightarrow\infty}m^{-1}R_{m}\left(
t\right)  >0$, e.g., when $\lim_{m\rightarrow\infty}\max_{i\in I_{0,m}}%
\sigma_{ii}<\infty$, i.e., when no $z_{i},i\in I_{0,m}$ eventually becomes
totally uninformative, and that $\liminf_{m\rightarrow\infty}\pi_{0,m}>0$
simply requires that there is always a positive proportion of true nulls.

Here we complement \autoref{prop:SLLNMarginal} by the following

\begin{corollary}
\label{prop:SLLNComp}For $\left\{  z_{i}\right\}  _{i=1}^{m}$, consider their
left-tailed p-values $\left\{  p_{i}\right\}  _{i=1}^{m}$ if $\Theta
_{0}=\left\{  0\right\}  $ or p-values $\left\{  p_{i}\right\}  _{i=1}^{m}$
defined by (\ref{defPvalComp}) if $\Theta_{0}$ is a composite null. If each
pair $\left(  z_{i},z_{j}\right)  $ with $i\neq j$ is bivariate Normal, then
(\ref{IneqA}) holds. If in addition (\ref{24}) holds, then (\ref{SLLNAB})
holds. If further%
\begin{equation}
\liminf_{m\rightarrow\infty}\pi_{0,m}>0\text{ \ and }\lim_{m\rightarrow\infty
}\min_{i\in I_{0,m}}\mathbb{P}\left(  p_{i}\leq\tau\right)  >0,
\label{condNullPval}%
\end{equation}
then (\ref{SLLNC}) holds.
\end{corollary}

\autoref{prop:SLLNComp} can potentially be used to consistently estimate the
FDR of an adaptive single-step MTP for testing composite nulls. The condition
$\lim_{m\rightarrow\infty}\min_{i\in I_{0,m}}\mathbb{P}\left(  p_{i}\leq
\tau\right)  >0$ holds when $\Theta_{0}=\left\{  0\right\}  $ and the
variances of $z_{i}$'s are uniformly bounded away from $0$ and $\infty$ or
when $p_{i},i\in I_{0,m}$ are identically distributed (as, e.g., in the case
where the classic definition of p-value is taken). Note that if $\inf_{m\geq
1}\sigma_{\left(  1\right)  }=0$ does happen, then some $z_{i}$'s will
eventually be equal to their means almost surely, and we will not make any
mistake testing these means and can test them separately. Further, condition
(\ref{condNullPval}) is to ensure $\liminf_{m\rightarrow\infty}m^{-1}%
R_{m}\left(  t\right)  >0$, which then yields (\ref{SLLNC}) from (\ref{SLLNAB}).

\subsection{Adaptive single-step multiple testing and consistent FDR
estimation}

\label{ConstFDREst}

Even though \autoref{prop:SLLNMarginal} and \autoref{prop:SLLNComp}
characterize, for multiple testing Normal means, a large class of dependence
under which the FDP of a single-step MTP concentrates almost surely around its
FDR, we can potentially improve the MTP\ as follows: firstly, we will make it
into an adaptive procedure by incorporating into it a consistent estimator of
the proportion $\pi_{0,m}$; secondly, using \autoref{prop:SLLNMarginal} and
\autoref{prop:SLLNComp} we can consistently estimate the FDR\ of the adaptive
MTP; finally, we can choose the rejection threshold for the adaptive MTP, so
that its FDR\ is as close as possible to a prespecified level as the number of
hypotheses to test increases.

Recall $z_{i}\sim\mathcal{N}_{1}\left(  \mu_{i},\sigma_{ii}\right)  $ and the
non-adaptive single-step MTP defined in \autoref{secPCSDef}. With an estimator
$\hat{\pi}_{1,m}\in\left[  0,1\right]  $ of $\pi_{1,m}$, at a nominal FDR
level $\alpha\in\left(  0,1\right)  $ we can construct an adaptive single-step
MTP as follows:\ define
\begin{equation}
\alpha_{m}\left(  \tau\right)  =\left(  1-\hat{\pi}_{1,m}\right)
\frac{\mathbb{P}\left(  p_{i_{0}}\leq\tau\right)  }{m^{-1}R_{m}\left(
\tau\right)  }1_{\left\{  R_{m}\left(  \tau\right)  \neq0\right\}  }
\label{eq6}%
\end{equation}
for some $i_{0}\in I_{0,m}$ and%
\begin{equation}
\hat{\tau}=\sup\left\{  \tau\in\left[  0,1\right]  :\alpha_{m}\left(
\tau\right)  \leq\alpha\right\}  , \label{eq22}%
\end{equation}
and reject $H_{i0}:\mu_{i}\in\Theta_{0}$ if and only if $p_{i}\leq\hat{\tau}$.
This adaptive procedure can be regarded as a slight extension of the procedure
of \cite{Storey:2004}. Note that $1-\hat{\pi}_{1,m}$ is an estimate of
$\pi_{0,m}$, and that the non-adaptive MTP is equivalent to setting $\hat{\pi
}_{1,m}=0$ in (\ref{eq6}). Since the rejection threshold $\hat{\tau}$ in
(\ref{eq22}) is almost surely at least as large as that obtained by setting
$\hat{\pi}_{1,m}=0$, the adaptive MTP can potentially reject more null
hypotheses and hence can be more powerful than the non-adaptive MTP at the
same nominal FDR level. However, when $\pi_{1,m}$ is very close to zero, there
may be little or no gain in power when using the adaptive MTP compared to its
non-adaptive counterpart. Further, the FDP of the adaptive MTP may not be
stable enough as an estimator of its FDR when the number of hypotheses to test
is small.

Denote the FDR of the adaptive MTP by $q$, i.e., $q=\mathbb{E}\left[
\mathrm{FDP}_{m}\left(  \hat{\tau}\right)  \right]  $. We have

\begin{theorem}
\label{ThmAdaptiveEstFDR} $\alpha_{m}\left(  \hat{\tau}\right)  \leq\alpha$
almost surely. Further, if the following conditions hold:

\begin{enumerate}
\item Each pair $\left(  z_{i},z_{j}\right)  ,i\neq j$ is bivariate Normal
such that (\ref{24}) holds and that $\lim_{m\rightarrow\infty}\mathbb{E}%
\left[  m^{-1}R_{m}\left(  t\right)  \right]  =Q\left(  t\right)  $ for a
continuous function $Q$;

\item $\left\{  p_{i}\right\}  _{i\in I_{0,m}}$ are identically distributed
and $\mathbb{P}\left(  p_{i_{0}}\leq t\right)  $ is continuous in $t\in\left[
0,1\right]  $;

\item $\lim_{m\rightarrow\infty}\pi_{1,m}=\pi_{1}$, $\lim_{t\downarrow
0}A\left(  t\right)  \neq\alpha$ and $\frac{d}{dt}A\left(  \tau_{1}\right)  $
exists and is nonzero, where%
\begin{equation}
A\left(  t\right)  =\frac{\left(  1-\pi_{1}\right)  \mathbb{P}\left(
p_{i_{0}}\leq\tau\right)  }{Q\left(  t\right)  }\ \ \text{and \ }\ \tau
_{1}=\sup\left\{  t\in\left[  0,1\right]  :A\left(  t\right)  \leq
\alpha\right\}  ; \label{defOracleProc}%
\end{equation}

\item $\lim_{m\rightarrow\infty}\mathbb{P}\left(  \left\vert \hat{\pi}%
_{1,m}-\pi_{1,m}\right\vert \rightarrow0\right)  =1$ and%
\begin{equation}
\lim_{m\rightarrow\infty}\mathbb{P}\left(  \min_{1\leq\tilde{m}\leq m}%
\tilde{m}^{-1}R_{\tilde{m}}\left(  \hat{\tau}\right)  >0\right)  =1,
\label{CondLowerLimR}%
\end{equation}

\end{enumerate}
then%
\begin{equation}
\lim_{m\rightarrow\infty}\mathbb{P}\left(  \hat{\tau}\rightarrow\tau
_{1}\right)  =1, \label{ThresholdCont}%
\end{equation}
and
\begin{equation}
\lim_{m\rightarrow\infty}\mathbb{P}\left(  \left\vert \mathrm{FDP}_{m}\left(
\hat{\tau}\right)  -q\right\vert \rightarrow0\right)  =1=\lim_{m\rightarrow
\infty}\mathbb{P}\left(  \left\vert \alpha_{m}\left(  \hat{\tau}\right)
-q\right\vert \rightarrow0\right)  , \label{FDPWLLN}%
\end{equation}
and $\lim_{m\rightarrow\infty}\mathrm{FDP}_{m}\left(  \hat{\tau}\right)
=A\left(  \tau_{1}\right)  $.
\end{theorem}

\autoref{ThmAdaptiveEstFDR} asserts that when the Normal random variables have
a PCS, $\hat{\pi}_{1,m}$ consistently estimates $\pi_{1,m}$ and some other
mild conditions are satisfied, $\alpha_{m}\left(  \hat{\tau}\right)  $
consistently estimates the FDR of the adaptive procedure and the FDP of the
procedure satisfies the weak law of large numbers. When the Normal random
variables all have unit variance, the adaptive MTP can be easily implemented
once we are able to verify $m^{-2}\left\Vert \mathbf{R}\right\Vert
_{1}=O\left(  m^{-\beta}\right)  $ for some\ $\beta>0$ without consistently
estimating $\mathbf{R}$. This is an advantage of FDR estimation under PCS.

The conditions on the function $A\left(  t\right)  $ defined by
(\ref{defOracleProc}), which essentially is the limiting process of the
process $\alpha_{m}\left(  t\right)  $, are standard, as already have been
(equivalently) used by, e.g., \cite{Storey:2004} and \cite{Hu:2010}.
\autoref{ThmAdaptiveEstFDR} can be regarded as a refinement of Theorem 4 of
\cite{Storey:2004} and of Theorem 4 of \cite{Hu:2010} with a consistent rather
than conservative estimator $\hat{\pi}_{0,m}$ of $\pi_{0,m}$ and as an upgrade
of these theorems to multiple testing one-sided nulls. The assertion
(\ref{ThresholdCont}), analogous to Theorem 5 of \cite{Storey:2004}, can be
regarded as an example of the continuity of a \textquotedblleft thresholding
functional\textquotedblright\ from the space of CDFs under the uniform
topology to the space of random variables under weak topology, originally
proved by \cite{Donoho:2006} for adaptive multiple testing means of
independent exponential random variables for the thresholding functional of
the BH procedure, for multiple testing the means of Normal random variables
under PCS via an adaptive single-step MTP. The condition (\ref{CondLowerLimR})
simply requires that the adaptive procedure always makes a positive proportion
of rejections and is needed to ensure (\ref{FDPWLLN}).

We remark that the condition \textquotedblleft$\left\{  p_{i}\right\}  _{i\in
I_{0,m}}$ are identically distributed\textquotedblright\ in
\autoref{ThmAdaptiveEstFDR} ensures%
\begin{equation}
\lim_{m_{0}\rightarrow\infty}m_{0}^{-1}V_{m}\left(  t\right)  =\mathbb{P}%
\left(  p_{i_{0}}\leq t\right)  \text{ almost surely for any }i_{0}\in
I_{0,m}\text{ and }t\in\left[  0,1\right]  , \label{eq33}%
\end{equation}
so that we can obtain (\ref{ThresholdCont}) and (\ref{FDPWLLN}). However, when
$\Theta_{0}$ is a bounded null, (\ref{eq33}) may not hold. In this case, we
can set%
\begin{equation}
\tilde{\alpha}_{m}\left(  \tau\right)  =\left(  1-\hat{\pi}_{1,m}\right)
\frac{\max_{1\leq i\leq m}\mathbb{P}\left(  p_{i}\leq\tau\right)  }%
{m^{-1}R_{m}\left(  \tau\right)  }1_{\left\{  R_{m}\left(  \tau\right)
\neq0\right\}  }, \label{eq33a}%
\end{equation}
define the rejection threshold
\[
\tilde{\tau}=\sup\left\{  \tau\in\left[  0,1\right]  :\tilde{\alpha}%
_{m}\left(  \tau\right)  \leq\alpha\right\}  ,
\]
and reject $H_{i0}:\mu_{i}\in\Theta_{0}$ if and only if $p_{i}\leq\tilde{\tau
}$. Even though doing so gives an adaptive single-step MTP, its power may
already have been compromised by the term $\max_{1\leq i\leq m}\mathbb{P}%
\left(  p_{i}\leq\tau\right)  $ in (\ref{eq33a}), which we will not
investigate here.

\section{Consistent proportion estimators under weak dependence}

\label{secAdaptiveTesting}

With the framework and key results laid out in
\autoref{secModel} for adaptive single-step multiple testing Normal means, we
can now move to proportion estimation under weak dependence. We will formally state the strategy for proportion estimation and proportion
estimators of \cite{Chen:2018a} and \cite{Chen:2019a} in \autoref{SecStrategy} and then show their
uniform consistency for each of the three types of nulls, i.e., the point
null $\Theta_{0}=\left\{  0\right\}  $, one-sided null $\Theta_{0}=\left(
-\infty,0\right)  $ and bounded null $\Theta_{0}=\left(  a,b\right)  $, for
weakly dependent Normal random variables. Specifically, the latter is done in
\autoref{secEstimator} for the point null, in
\autoref{SecBoundedNullDep} for the bounded null, and in
\autoref{secOneSidedNullDep} for the one-sided null. These
proportion estimators can be used by the adaptive single-step MTPs introduced
in \autoref{ConstFDREst}.

\subsection{Construction of proportion estimators via integral equations}

\label{SecStrategy}

Let $\boldsymbol{\mu}=\left(  \mu_{1},...,\mu_{m}\right)  ^{\top}$ and recall
$\pi_{1,m}$ defined by (\ref{defPiOne}). Denote by $F_{\mu_{i}}$ the CDF of
$z_{i}$ for $i\in\left\{  1,\ldots,m\right\}  $, and suppose each $F_{\mu_{i}%
}$ is a member of a set $\mathcal{F}$ of CDFs such that $\mathcal{F}=\left\{
F_{\mu}:\mu\in U\right\}  $ for some non-empty, interval $U$ in $\mathbb{R}$
that has a non-empty interior. Note that $U=\mathbb{R}$ for Normal random variables. Uniformly consistent estimators of $\pi_{1,m}$
for each of the three types of nulls have been constructed by
\cite{Chen:2018a} and \cite{Chen:2019a} using solutions to Lebesgue-Stieltjes
integral equations. Specifically, each such estimator is an \textquotedblleft
empirical matching function\textquotedblright%
\begin{equation}
\hat{\varphi}_{m}\left(  t,\mathbf{z}\right)  =m^{-1}\sum_{i=1}^{m}\left\{
1-K\left(  t,z_{i}\right)  \right\}  , \label{eq4b}%
\end{equation}
where $K:\mathbb{R}^{2}\rightarrow\mathbb{R}$ is a \textquotedblleft matching
function\textquotedblright\ that does not depend on any $\mu\in\Theta_{1}$ and
satisfies the Lebesgue-Stieltjes integral equation%
\[
\psi\left(  t,\mu\right)  =\int K\left(  t,x\right)  dF_{\mu}\left(  x\right)
\text{ for all }\mu\in U
\]
for a \textquotedblleft discriminant function\textquotedblright\ $\psi$ such
that $\lim_{t\rightarrow\infty}\psi\left(  t,\mu\right)  =1$ for $\mu\in
\Theta_{0}$ and $\lim_{t\rightarrow\infty}\psi\left(  t,\mu\right)  =0$ for
$\mu\in\Theta_{1}$. Further, $\mathbb{E}\left\{  \hat{\varphi}_{m}\left(
t,\mathbf{z}\right)  \right\}  =\varphi_{m}\left(  t,\boldsymbol{\mu}\right)
$, where%
\begin{equation}
\varphi_{m}\left(  t,\boldsymbol{\mu}\right)  =m^{-1}\sum_{i=1}^{m}\left\{
1-\psi\left(  t,\mu_{i}\right)  \right\}  \label{eq4a}%
\end{equation}
is the \textquotedblleft average discriminant function\textquotedblright, and
the \textquotedblleft Oracle\textquotedblright\ $\Lambda_{m}\left(
\boldsymbol{\mu}\right)  =\lim_{t\rightarrow\infty}\varphi_{m}\left(
t,\boldsymbol{\mu}\right)  =\pi_{1,m}$ for any fixed $m$ and $\boldsymbol{\mu
}$. We will reserve the notation $\left(  \psi,K\right)  $ for a pair of
discriminant function and matching function and the notations $\varphi
_{m}\left(  t,\boldsymbol{\mu}\right)  $ and $\hat{\varphi}_{m}\left(
t,\mathbf{z}\right)  $ as per (\ref{eq4a}) and (\ref{eq4b}) unless otherwise
noted. We refer interested readers to \cite{Chen:2018a} and \cite{Chen:2019a}
for more details on how these estimators are constructed, their advantages
over other methods of constructing proportion estimators, and discussions on
several existing, leading proportion estimators, and specifically to
\cite{Jin:2008} for the origin of the method of construction presented above.

Let $r_{\mu}$ be the modulus of the characteristic function (CF) $\hat{F}%
_{\mu}$ of the CDF$\ F_{\mu}$ of $X\sim\mathcal{N}_{1}\left(  \mu,\sigma
^{2}\right)  $ for $\mu\in U$, i.e., $r_{\mu}\left(  t\right)  =\exp\left(
-2^{-1}t^{2}\sigma^{2}\right)  $, which is independent of $\mu$. The
constructions and proportion estimators (hereafter referred to as
\textquotedblleft proposed estimators\textquotedblright) provided by
\cite{Chen:2018a} and \cite{Chen:2019a} for Normal random variables are
restated by the following theorem:

\begin{theorem}
\label{ThmAllConstructions}Let $\omega$ be non-negative, even and of bounded
variation and Lebesgue integrate to $1$ on $\left[  -1,1\right]  $.

\begin{enumerate}
\item For the point null\ $\Theta_{0}=\left\{  0\right\}  $, $\left(
\psi,K\right)  =\left(  \psi_{1,0}\left(  t,\mu;0\right)  ,K_{1,0}\left(
t,x;0\right)  \right)  $, where for any $\mu^{\prime}\in U$,\
\[
\left\{
\begin{array}
[c]{l}%
K_{1,0}\left(  t,x;\mu^{\prime}\right)  ={\int_{\left[  -1,1\right]  }}%
\dfrac{\omega\left(  s\right)  \cos\left\{  ts\left(  x-\mu^{\prime}\right)
\right\}  }{r_{\mu^{\prime}}\left(  ts\right)  }ds\\
\psi_{1,0}\left(  t,\mu;\mu^{\prime}\right)  =\int K_{1,0}\left(
t,x;\mu^{\prime}\right)  dF_{\mu}\left(  x\right)  ={\int_{\left[
-1,1\right]  }}\omega\left(  s\right)  \cos\left\{  ts\left(  \mu-\mu^{\prime
}\right)  \right\}  ds
\end{array}
\right.  .
\]

\item For the one-sided null $\Theta_{0}=\left(  -\infty,0\right)  $, $\left(
\psi,K\right)  $ is%
\begin{equation}
\left\{
\begin{array}
[c]{l}%
K\left(  t,x\right)  =2^{-1}-K_{1}\left(  t,x\right)  -2^{-1}K_{1,0}\left(
t,x;0\right) \\
\psi\left(  t,\mu\right)  =2^{-1}-\psi_{1}\left(  t,\mu\right)  -2^{-1}%
\psi_{1,0}\left(  t,\mu;0\right)
\end{array}
\right.  , \label{V-a}%
\end{equation}
where
\[
K_{1}\left(  t,x\right)  =\frac{1}{2\pi}\int_{0}^{1}dy\int_{-1}^{1}\left[
\frac{\sin\left(  ytsx\right)  }{y}\left\{  \frac{d}{ds}\frac{1}{r_{0}\left(
tys\right)  }\right\}  +\frac{tx\cos\left(  tysx\right)  }{r_{0}\left(
tys\right)  }\right]  ds,
\]
and%
\[
\psi_{1}\left(  t,\mu\right)  =\int K_{1}\left(  t,x\right)  dF_{\mu}\left(
x\right)  =\frac{1}{2\pi}\int_{0}^{t}dy\int_{-1}^{1}\mu\exp\left(  \iota
ys\mu\right)  ds.
\]

\item For the bounded null $\Theta_{0}=\left(  a,b\right)  $, $\left(
\psi,K\right)  $ is%
\begin{equation}
\left\{
\begin{array}
[c]{l}%
K\left(  t,x\right)  =K_{1}\left(  t,x\right)  -2^{-1}\left\{  K_{1,0}\left(
t,x;a\right)  +K_{1,0}\left(  t,x;b\right)  \right\} \\
\psi\left(  t,\mu\right)  =\psi_{1}\left(  t,\mu\right)  -2^{-1}\left\{
\psi_{1,0}\left(  t,\mu;a\right)  +\psi_{1,0}\left(  t,\mu;b\right)  \right\}
\end{array}
\right.  , \label{IV-a}%
\end{equation}
where
\begin{equation}
K_{1}\left(  t,x\right)  =\frac{t}{2\pi}\int_{a}^{b}dy\int_{\left[
-1,1\right]  }\frac{\cos\left\{  ts\left(  x-y\right)  \right\}  }%
{r_{0}\left(  ts\right)  }ds \label{eq13e}%
\end{equation}
and
\[
\psi_{1}\left(  t,\mu\right)  =\int K_{1}\left(  t,x\right)  dF_{\mu}\left(
x\right)  =\frac{1}{\pi}\int_{\left(  \mu-b\right)  t}^{\left(  \mu-a\right)
t}\frac{\sin y}{y}dy.
\]

\end{enumerate}
\end{theorem}

When the difference
\begin{equation}
e_{m}\left(  t\right)  =\hat{\varphi}_{m}\left(  t,\mathbf{z}\right)
-\varphi_{m}\left(  t,\boldsymbol{\mu}\right)  \label{eq2d}%
\end{equation}
is small for large $t$, $\hat{\varphi}_{m}\left(  t,\mathbf{z}\right)  $ will
accurately estimate $\pi_{1,m}$. Since $\varphi_{m}\left(  t,\boldsymbol{\mu
}\right)  =\pi_{1,m}$ or $\hat{\varphi}_{m}\left(  t,\mathbf{z}\right)
=\pi_{1,m}$ rarely happens, $\hat{\varphi}_{m}\left(  t,\mathbf{z}\right)  $
often employs an increasing sequence $\left\{  t_{m}\right\}  _{m\geq1}$ with
$\lim_{m\rightarrow\infty}t_{m}=\infty$ in order to achieve \textquotedblleft
consistency\textquotedblright\ in the sense that%
\begin{equation}
\Pr\left\{  \left\vert \pi_{1,m}^{-1}\hat{\varphi}_{m}\left(  t_{m}%
,\mathbf{z}\right)  -1\right\vert \rightarrow0\right\}  \rightarrow
1\text{\quad as \quad}m\rightarrow\infty. \label{defConsistency}%
\end{equation}
We refer to $t_{m}$ as the \textquotedblleft speed of
convergence\textquotedblright\ of $\hat{\varphi}_{m}\left(  t_{m}%
,\mathbf{z}\right)  $ as did by \cite{Chen:2018a} and \cite{Chen:2019a}. By
duality, $\varphi_{m}^{\ast}\left(  t,\boldsymbol{\mu}\right)  =1-\varphi
_{m}\left(  t,\boldsymbol{\mu}\right)  $ satisfies $\pi_{0,m}=\lim
_{t\rightarrow\infty}\varphi_{m}^{\ast}\left(  t,\boldsymbol{\mu}\right)  $
for any fixed $m$ and $\boldsymbol{\mu}$. Accordingly, $\hat{\varphi}_{m}^{\ast}\left(
t,\mathbf{z}\right)  =1-\hat{\varphi}_{m}\left(  t,\mathbf{z}\right)  $, as an
estimator of $\pi_{0,m}$, satisfies $\mathbb{E}\left\{  \hat{\varphi}%
_{m}^{\ast}\left(  t,\mathbf{z}\right)  \right\}  =\varphi_{m}^{\ast}\left(
t,\boldsymbol{\mu}\right)  $ for any fixed $m,t$ and $\boldsymbol{\mu}$, and the oscillation of
$\hat{\varphi}_{m}^{\ast}\left(t,\mathbf{z}\right) $ is also $e_{m}\left(  t\right)$.

\subsection{Consistency of proportion estimator for the case of a point null}

\label{secEstimator}

For the point null $\Theta_{0}=\left\{  0\right\}  $,\ $K_{1,0}$ and
$\psi_{1,0}$ in \autoref{ThmAllConstructions} become
\begin{equation}
\left\{
\begin{array}
[c]{l}%
K_{1,0}\left(  t,x;0\right)  ={\int_{\left[  -1,1\right]  }}\omega\left(
s\right)  \exp\left(  2^{-1}t^{2}s^{2}\sigma^{2}\right)  \cos\left(
tsx\right)  ds\\
\psi\left(  t,\mu\right)  =\psi_{1,0}\left(  t,\mu;0\right)  ={\int_{\left[
-1,1\right]  }}\omega\left(  s\right)  \cos\left(  ts\mu\right)  ds
\end{array}
\right.  . \label{eqb14}%
\end{equation}
Write $K_{1,0}\left(  t,x;0\right)  $ as $G_{\sigma}\left(  t,x\right)  $ and
let%
\[
\phi_{\mu,\sigma}\left(  x\right)  =\left(  \sqrt{2\pi}\sigma\right)
^{-1}\exp\left(  -2^{-1}\sigma^{-2}\left(  x-\mu\right)  ^{2}\right)  \text{
for }\mu\in\mathbb{R},\sigma>0.
\]
Then $\mathbb{E}\left[  G_{\sigma}\left(  t,X\right)  \right]  =\psi\left(
t,\mu\right)  =\hat{\omega}\left(  t\mu\right)  $ for all $t$ and $\mu$ when
$X\sim\mathcal{N}_{1}\left(  \mu,\sigma^{2}\right)  $, where $\hat{\omega}$
denotes the Fourier transform of $\omega$, and the identity
\begin{equation}
\left(  G_{1}\left(  t;\cdot\right)  \ast\phi_{0,1}\right)  \left(
\mu\right)  =\psi\left(  t,\mu\right)  =\left(  G_{\sigma}\left(
t;\cdot\right)  \ast\phi_{0,\sigma}\right)  \left(  \mu\right)  \label{eq25}%
\end{equation}
holds, where $\ast$ denotes convolution. We remark that $K_{1,0}$ in
(\ref{eqb14}) and $\hat{\varphi}_{m}$ in (\ref{eq4b}) with this $K_{1,0}$
generalize corresponding functions in \cite{Jin:2008} by allowing each $z_{i}$
to have its own variance $\sigma_{ii}$. The estimator $\hat{\varphi}%
_{m}\left(  t,\mathbf{z}\right)  $ based on this $\left(  \psi,K\right)  $
reduces to Jin's estimator in \cite{Jin:2008} when $\sigma_{ii}=1$ for all $i$.

Recall (\ref{eq23}), i.e., $m^{-2}\left\Vert \mathbf{\Sigma}\right\Vert
_{1}=O\left(  m^{-\delta}\right)  $ for some $\delta>0$ and $\sigma_{\left(
m\right)  }=\max_{1\leq i\leq m}\sigma_{ii}$ and set $\left\Vert
\omega\right\Vert _{\infty}=\sup_{s\in\left[  -1,1\right]  }\left\vert
\omega\left(  s\right)  \right\vert $. We have

\begin{theorem}
\label{LmBoundDiffPhaseFuncWTumix}Let $p_{m}\left(  t\right)  =m^{-1}%
+2t^{2}m^{-2}\left\Vert \mathbf{\Sigma}\right\Vert _{1}$ and assume each pair
$\left(  z_{i},z_{j}\right)  ,i\neq j$ is bivariate Normal. Then%
\begin{equation}
\mathbb{E}\left[  e_{m}^{2}\left(  t\right)  \right]  \leq4\left\Vert
\omega\right\Vert _{\infty}^{2}\exp\left(  t^{2}\sigma_{\left(  m\right)
}\right)  p_{m}\left(  t\right)  , \label{eqx3}%
\end{equation}
and, for any $\varepsilon>0$, with probability at least $1-\varepsilon
^{-2}p_{m}\left(  t\right)  $,%
\begin{equation}
\left\vert e_{m}\left(  t\right)  \right\vert \leq2\varepsilon\left\Vert
\omega\right\Vert _{\infty}\exp\left(  2^{-1}t^{2}\sigma_{\left(  m\right)
}\right)  . \label{eq14}%
\end{equation}
If $\inf_{m\geq1}\sigma_{\left(  m\right)  }>0$ and (\ref{eq23}) holds, then,
for any three constants $\eta$, $\gamma$ and $\delta^{\prime}$ such that
$0<\gamma<\eta$ and $2\eta<\delta^{\prime}<\min\left\{  \delta,1\right\}  $,
and $t_{\gamma,m}=\sqrt{2\sigma_{\left(  m\right)  }^{-1}\gamma\ln m}$,
\begin{equation}
\sup\nolimits_{t\in\left(  0,t_{\gamma,m}\right]  }\left\vert e_{m}\left(
t\right)  \right\vert \leq2\left\Vert \omega\right\Vert _{\infty}%
m^{\gamma-\eta}, \label{eqb97}%
\end{equation}
holds with probability at least $1-Cm^{-\left(  \delta^{\prime}-2\eta\right)
}$ for some constant $C$ that is determined by (\ref{eq23}).
\end{theorem}

\autoref{LmBoundDiffPhaseFuncWTumix} provides an integrated view on the
relationship between the strength of dependence among the Normal random
variables $m^{-2}\left\Vert \mathbf{\Sigma}\right\Vert _{1}$, their maximal
variance $\sigma_{\left(  m\right)  }$, and the magnitude of oscillations of
$e_{m}\left(  t\right)  $. In particular, it asserts that, when the Normal
random variables are weakly dependent, with overwhelming probability, for any
$\gamma$ and $\eta$ such that $0<\gamma<\eta$ and $2\eta<\min\left\{
\delta,1\right\}  $, the stochastic fluctuation of $\left\vert e_{m}\left(
t\right)  \right\vert $ is upper bounded in order by $m^{\gamma-\eta}$
uniformly for $t\in\left(  0,t_{\gamma,m}\right]  $. It also provides in
(\ref{eqx3}) an upper bound on the variance of the estimator $\hat{\varphi
}_{m}\left(  t,\mathbf{z}\right)  $, which helps characterize its stability.
The condition $\inf_{m\geq1}\sigma_{\left(  m\right)  }>0$ requires that not
all Normal random variables are equal to their means almost surely for all $m$
and is easily satisfied. In case $\sigma_{\left(  \tilde{m}\right)  }=0$ for
some positive integer $\tilde{m}$, then $\pi_{1,\tilde{m}}=\tilde{m}^{-1}%
\sum_{i=1}^{\tilde{m}}1_{\left\{  \tilde{z}_{i} \in\Theta_{1}\right\}  }$
almost surely and is known, where $\tilde{z}_{i}$ is a realization of $z_{i}$
for $i=1,\cdots,\tilde{m}$.

When the Normal random variables are mutually independent and have unit
variance, $m^{-2}\left\Vert \mathbf{\Sigma}\right\Vert _{1}=m^{-1}$,
$\delta=1$ and $\sigma_{\left(  m\right)  }=1$ in
\autoref{LmBoundDiffPhaseFuncWTumix}. Accordingly, the range for $\gamma$ is
$\left(  0,0.5\right)  $ in \autoref{LmBoundDiffPhaseFuncWTumix} whereas the
range for $\gamma$ is $(0,0.5]$ in \cite{Jin:2008} and \cite{Chen:2018a},
meaning that the difference between the two is the singleton $\left\{
0.5\right\}  $. Such a difference is mainly due to Lemmas 4 and 5 in
\cite{Jin:2008}, no longer applicable to the dependence setting here, that
give
\begin{equation}
\sup_{t\in(\left.  0,\sqrt{2\gamma\ln m}\right]  }\left\vert e_{m}\left(
t\right)  \right\vert =O\left(  \frac{m^{\gamma-1/2}}{\sqrt{\ln m}}\right)
\text{ \ for \ }\gamma\in\left( 0,0.5\right] , \label{eq24}%
\end{equation}
where the gain $\left(  \ln m\right)  ^{-1/2}$ in (\ref{eq24}) allows
$\gamma=0.5$. Nonetheless, this does not reduce the applicability of the
proportion estimator.

Using the bound on $e_{m}\left(  t\right)  $ provided in
\autoref{LmBoundDiffPhaseFuncWTumix}, we obtain uniform consistency of the
estimator as

\begin{theorem}
\label{ThmConsistPlugEst}Assume each pair $\left(  z_{i},z_{j}\right)  ,i\neq
j$ is bivariate Normal such that $\inf_{m\geq1}\sigma_{\left(  m\right)  }>0$.
Let $\tilde{u}_{m}=\min_{\left\{  j:\mu_{j}\neq0\right\}  }\left\vert \mu
_{j}\right\vert $. Further, assume (\ref{eq23}), $\lim_{m\rightarrow\infty
}\tilde{u}_{m}\sqrt{2\gamma\ln m}=\infty$ and\ $\lim_{m\rightarrow\infty}%
\pi_{1,m}m^{\eta-\gamma}=\infty$, where $\gamma$ and $\eta$ are specified in
\autoref{LmBoundDiffPhaseFuncWTumix}. Then, with $t_{\gamma,m}$ defined by
\autoref{LmBoundDiffPhaseFuncWTumix},%
\begin{equation}
\mathbb{P}\left(  \left\vert \pi_{1,m}^{-1}\hat{\varphi}_{m}\left(
t_{\gamma,m},\mathbf{z}\right)  -1\right\vert \rightarrow0\right)
\rightarrow1\text{ \ as \ }m\rightarrow\infty. \label{eq20}%
\end{equation}

\end{theorem}

\autoref{ThmConsistPlugEst} justifies the consistency of the estimator for a
range of $\pi_{1,m}$ values when $z_{i}$'s are weakly dependent, going beyond
the settings of strongly mixing and short-range dependence that have been
covered by \cite{Jin:2007} and \cite{Jin:2008} for Jin's estimator. In
particular, it implies the consistency of Jin's estimator of \cite{Jin:2008}
when $z_{i}$'s have unit variance and are mutually independent. The condition
$\lim_{m\rightarrow\infty}\tilde{u}_{m}\sqrt{2\gamma\ln m}=\infty$ simply
requires the minimal magnitude of the nonzero Normal means not to be extremely
small, and it is satisfied when $\tilde{u}_{m}\geq\ln\ln m/\sqrt{\ln m}$ as
assumed by \cite{Jin:2008} and \cite{Chen:2018a}. Note that since we already
have the variance upper bound for $e_{m}\left(  t\right)  $ in (\ref{eqx3}),
we can choose a speed of convergence $t_{m}$ different than $t_{\gamma,m}$ for
$t$ in $\hat{\varphi}_{m}\left(  t,\mathbf{z}\right)  $, so that
(\ref{defConsistency}) still holds for different ranges of $\pi_{1,m}$.

In \autoref{ThmConsistPlugEst}, the condition $\pi_{1,m}m^{\eta-\gamma
}\rightarrow\infty$ means that, under weak dependence, the sparsest $\pi
_{1,m}$ that can be consistently estimated by the estimator has to be denser
than $m^{\gamma-\eta}$ with $0<\gamma<\eta$ and $2\eta<\min\left\{
\delta,1\right\}  $. This is due to the fact that the stochastic fluctuation
of $\hat{\varphi}_{m}\left(  t_{\gamma,m},\mathbf{z}\right)  $ around
$\varphi_{m}\left(  t,\boldsymbol{\mu}\right)  $ under weak dependence, even
though converging to $0$ as $m\rightarrow\infty$, can be as large in order as
$m^{\gamma-\eta}$. In contrast, when the Normal random variables are
independent and have unit variance, the sparsest $\pi_{1,m}$ in
\cite{Jin:2008} and \cite{Chen:2018a} is allowed to be $m^{\gamma-1/2}$ with
$\gamma\in(0,0.5]$ due to (\ref{eq24}). In other words, the speed of
convergence of $\hat{\varphi}_{m}\left(  t_{\gamma,m},\mathbf{z}\right)  $
should not \textquotedblleft override\textquotedblright\ the index $\delta$ of
weak dependence. Regardless, the level of sparsity of $\pi_{1,m}$ for which
$\pi_{1,m}$ can be consistently estimated by the estimator is largely
determined by the index $\delta$ of weak dependence in (\ref{eq23}).

The sequence $\left\{  t_{\gamma,m}\right\}  _{m\geq1}$ for $\hat{\varphi}%
_{m}\left(  t_{\gamma,m},\mathbf{z}\right)  $ is also referred to as the
\textquotedblleft thresholding sequence\textquotedblright\ for the estimator.
When the Normal random variables have unit variance, it is also the
thresholding sequence for Jin's estimator. When $\gamma$ is set to be the
supremum of its allowed value, i.e., when $\gamma=0.5\delta$ is set, where
$\delta$ is the index of weak dependence, we say that the corresponding
thresholding sequence is maximal. When the Normal random variables are
mutually independent and have unit variance, \cite{Jin:2008} and
\cite{Chen:2018a} provided strong evidence on the excellent performance of
$\varphi_{m}\left(  t_{m};\mathbf{z}\right)  $ when it uses the maximal
thresholding sequence $t_{0.5,m}=\sqrt{\ln m}$ for which $\gamma=0.5$.

\subsection{Consistency of proportion estimator for the case of a bounded
null}

\label{SecBoundedNullDep}

For the bounded null $\Theta_{0}=\left(  a,b\right)  $, we assume $\sigma
_{ii}=\sigma^{2}$ for some fixed variance $\sigma^{2}$ for all $i\in\left\{
1,\ldots,m\right\}  $ to avoid unnecessary notational burdens of dealing with
different $\sigma_{ii}$ without significant gains in sharper statistical
results. After all, when $\sigma_{ii}$'s are different but known, we can scale
$z_{i}$'s (so that the scaled $z_{i}$'s have unit variance) and adjust the
bounded null. For the bounded null $\Theta_{0}=\left(  a,b\right)  $, $K_{1}$
in \autoref{ThmAllConstructions} becomes
\[
K_{1}\left(  t,x\right)  =\frac{t}{2\pi}\int_{a}^{b}dy\int_{\left[
-1,1\right]  }\exp\left(  2^{-1}t^{2}s^{2}\sigma^{2}\right)  \cos\left\{
ts\left(  x-y\right)  \right\}  ds,
\]
and we have the following theorem on the uniform consistency of the estimator
$\hat{\varphi}_{m}\left(  t,\mathbf{z}\right)  $:

\begin{theorem}
\label{ThmIVaDep}Let $p_{m}\left(  t\right)  =m^{-1}+2t^{2}m^{-2}\left\Vert
\mathbf{\Sigma}\right\Vert _{1}$ and assume each pair $\left(  z_{i}%
,z_{j}\right)  ,i\neq j$ is bivariate Normal. Then
\begin{equation}
\mathbb{V}\left\{  e_{m}\left(  t\right)  \right\}  \leq2p_{m}\left(
t\right)  \frac{\exp\left(  t^{2}\sigma^{2}\right)  -1}{t^{2}\sigma^{2}%
}\left\{  \frac{t^{2}\left(  b-a\right)  ^{2}}{\pi^{2}}+16\left\Vert
\omega\right\Vert _{\infty}^{2}\right\}  . \label{eqx3cx}%
\end{equation}
Set $u_{m}=\min_{u\in\left\{  a,b\right\}  }\min_{\left\{  j:\mu_{j}\neq
u\right\}  }\left\vert \mu_{j}-u\right\vert $ and $t_{m}=\sqrt{\gamma
\sigma^{-2}\ln m}$ for some $\gamma>0$. If in addition
\begin{equation}
t_{m}^{-1}\left(  1+u_{m}^{-1}\right)  =o\left(  \pi_{1,m}\right)  \text{
\ \ and \ \ }m^{-2+\gamma}\left\Vert \mathbf{\Sigma}\right\Vert _{1}\ln
m=o\left(  \pi_{1,m}^{2}\right)  , \label{eqx3c}%
\end{equation}
then $\hat{\varphi}_{m}\left(  t_{m},\mathbf{z}\right)  $ is consistent.
\end{theorem}

From \autoref{ThmIVaDep}, we can easily obtain the uniform consistency of
$\hat{\varphi}_{m}\left(  t_{m},\mathbf{z}\right)  $ for a range of $\pi
_{1,m}$ and $\boldsymbol{\mu}$ that depends on the index $\delta$ of weak
dependence. When $\left\{  z_{i}\right\}  _{i=1}^{m}$ are independent, the
second condition in (\ref{eqx3c}) becomes $m^{\left( \gamma-1\right)  /2}%
\sqrt{\ln m}=o\left(  \pi_{1,m}\right)  $. In contrast, if $\left\{
z_{i}\right\}  _{i=1}^{m}$ are not independent such that $m^{-2}\left\Vert
\mathbf{\Sigma}\right\Vert _{1}=O\left(  m^{-\delta}\right)  $ for\ some
$\delta\in\left(  0,1\right)  $, then the second condition in (\ref{eqx3c})
becomes $m^{\left(  \gamma-\delta\right)  /2}\sqrt{\ln m}=o\left(  \pi
_{1,m}\right)  $. Accordingly, the estimator $\varphi_{m}\left(
t_{m};\mathbf{z}\right)  $ with $t_{m}=\sqrt{\gamma\sigma^{-2}\ln m}$ may not
be consistent when $\gamma\geq\delta$ whereas it can be when $\gamma<\delta$.
However, regardless of if $\left\{  z_{i}\right\}  _{i=1}^{m}$ are independent
or not, the sparsest $\pi_{1,m}$ it is able to consistently estimate at speed
$t_{m}=\sqrt{\gamma\sigma^{-2}\ln m}$ should be larger in order than
$\ln^{-1/2}m$, as suggested by (\ref{eqx3c}). We remark that, similar to the
case of a point null, the variance upper bound for $e_{m}\left(  t\right)  $
in (\ref{eqx3cx}) allows us to choose a speed of convergence $t_{m}$ different
than $\sqrt{\gamma\sigma^{-2}\ln m}$ for $t$ in $\hat{\varphi}_{m}\left(
t,\mathbf{z}\right)  $, so that (\ref{defConsistency}) still holds for
different ranges of $\pi_{1,m}$.

\subsection{Consistency of proportion estimator for the case of a one-sided
null}

\label{secOneSidedNullDep}

Recall the one-sided null $\Theta_{0}=\left(  -\infty,0\right)  $. In this
setting, $K_{1}$ in \autoref{ThmAllConstructions} becomes
\[
K_{1}\left(  t,x\right)  =\frac{1}{2\pi}\int_{0}^{1}dy\int_{-1}^{1}\exp\left(
2^{-1}t^{2}y^{2}s^{2}\sigma^{2}\right)  \left\{  syt^{2}\sigma^{2}\sin\left(
ytsx\right)  +tx\cos\left(  tysx\right)  \right\}  ds.
\]
For this null, we assume $\sigma_{ii}=\sigma^{2}$ for some fixed variance
$\sigma^{2}$ for all $i\in\left\{  1,\ldots,m\right\}  $ for the same reason
stated in \autoref{SecBoundedNullDep}. We have the following result on the
uniform consistency of the estimator $\hat{\varphi}_{m}\left(  t,\mathbf{z}%
\right)  $:

\begin{theorem}
\label{ThmTypeI-VDep}Let $c_{\ast}=\sup_{m\geq1}\max_{1\leq i<j\leq
m}\left\vert \sigma_{ij}\right\vert $ and $\mu_{\ast}=\max_{1\leq i\leq
m}\left\vert \mu_{i}\right\vert $. If $c_{\ast}<\sigma^{2}$ and each pair
$\left(  z_{i},z_{j}\right)  ,i\neq j$ is bivariate Normal, then%
\begin{equation}
\mathbb{V}\left\{  e_{m}\left(  t\right)  \right\}  \leq\frac{\exp\left(
t^{2}\sigma^{2}\right)  -1}{\pi^{2}}\left\{  p_{m}\left(  t\right)
t^{2}\sigma^{2}+\frac{p_{2,m}}{\sigma^{2}}\right\}  , \label{eq12y}%
\end{equation}
where $p_{m}\left(  t\right)  =m^{-1}+2t^{2}m^{-2}\left\Vert \mathbf{\Sigma
}\right\Vert _{1}$ and
\[
p_{2,m}=m^{-1}\left(  \sigma^{2}+\mu_{\ast}\right)  +\frac{4K_{0}^{2}%
\sigma^{2}\left(  \sigma+\mu_{\ast}\right)  ^{2}}{\left(  \sigma^{2}-c_{\ast
}\right)  }m^{-2}\left\Vert \boldsymbol{\Sigma}\right\Vert _{1}%
\]
for a constant $K_{0}>0$. Set $t_{m}=\sqrt{\gamma\sigma^{-2}\ln m}$ for some
$\gamma>0$. If in addition
\begin{equation}
t_{m}^{-1}\left(  1+\tilde{u}_{m}^{-1}\right)  =o\left(  \pi_{1,m}\right)
\text{ \ and \ }m^{-2+\gamma}\left\Vert \mathbf{\Sigma}\right\Vert _{1}\left(
\ln^{4}m+\mu_{\ast}^{2}+\mu_{\ast}\right)  =o\left(  \pi_{1,m}^{2}\right)  ,
\label{eq3xd}%
\end{equation}
then $\varphi_{m}\left(  t_{m};\mathbf{z}\right)  $ is consistent.
\end{theorem}

From \autoref{ThmTypeI-VDep}, we can easily obtain the uniform consistency of
$\hat{\varphi}_{m}\left(  t_{m},\mathbf{z}\right)  $ for a range of $\pi
_{1,m}$ and $\boldsymbol{\mu}$ that depends on the index $\delta$ of weak
dependence. If $\left\{  z_{i}\right\}  _{i=1}^{m}$ are independent,
$t_{m}=\sqrt{\gamma\sigma^{-2}\ln m}$ and $t_{m}^{-1}\left(  1+\tilde{u}%
_{m}^{-1}\right)  =o\left(  \pi_{1,m}\right)  $, then \autoref{ThmTypeI-VDep}
asserts that $\hat{\varphi}_{m}\left(  t_{m},\mathbf{z}\right)  $ is
consistent when $m^{\gamma-1}\left(  \ln^{4}m+\mu_{\ast}^{2}+\mu_{\ast
}\right)  =o\left(  \pi_{1,m}^{2}\right)  $. Note that we could have directly
used the uniform consistency class provided by Theorem 6 of \cite{Chen:2019a}
when $\left\{  z_{i}\right\}  _{i=1}^{m}$ are independent, which says that
$\hat{\varphi}_{m}\left(  t_{m},\mathbf{z}\right)  $ with $t_{m}=\sqrt
{\gamma\sigma^{-2}\ln m}$ is consistent when $t_{m}^{-1}\left(  1+\tilde
{u}_{m}^{-1}\right)  =o\left(  \pi_{1,m}\right)  $ and $0<\gamma
<\gamma^{\prime}<1$ and $\mu_{\ast}^{2}m^{\gamma^{\prime}-1}=o\left(
1\right)  $. Similar to the cases of a point null and a one-sided null, the
variance upper bound for $e_{m}\left(  t\right)  $ in (\ref{eq12y}) allows us
to choose a speed of convergence $t_{m}$ different than $\sqrt{\gamma
\sigma^{-2}\ln m}$ for $t$ in $\hat{\varphi}_{m}\left(  t,\mathbf{z}\right)
$, so that (\ref{defConsistency}) still holds for different ranges of
$\pi_{1,m}$.

\section{Simulation study}

\label{SecNumericalStudies}

In this section, we provide a simulation study on the proposed estimators, by
also comparing them with the \textquotedblleft MR\textquotedblright\ estimator
$\hat{\pi}_{\mathrm{MR}}$ of \cite{Meinshausen:2006}. We choose $\hat{\pi
}_{\mathrm{MR}}$ for a comparison with the proposed estimators since $\hat
{\pi}_{\mathrm{MR}}$ is the only theoretically proven consistent estimator of
$\pi_{1,m}$, besides those of \cite{Jin:2008}, \cite{Chen:2018a} and
\cite{Chen:2019a}, that is able to handle $\pi_{1,m}=m^{-\vartheta}$ for
$\vartheta\in\left[  0,0.5\right)  $ without the need to employ the Bayesian
two-component mixture model on $\left\{  z_{i}\right\}  _{i=1}^{p}$ or their
p-values. However, since $\hat{\pi}_{\mathrm{MR}}$ is based on p-values, we
will convert each observed Normal random variable $z_{i}$ into p-value $p_{i}$
and apply them to $\left\{  p_{i}\right\}  _{i=1}^{m}$ to estimate $\pi_{1,m}%
$. Note that the consistency of $\hat{\pi}_{\mathrm{MR}}$ requires that
p-values are identically distributed under the false nulls and that this is
not satisfied by the simulation design to be given next.

For one-side null,
we will also compare our estimator with a version, referred to as the ``HD'' estimator, of one minus the proportion estimator of Section 4.3 of \cite{Hoang:2022b}
that has both of its two tuning parameters as $0.5$ (since our previous study provided evidence that this estimator with its second tuning parameter (for constructing randomized p-values)
optimally determined from data performed worse than the ``HD'' estimator).
For other settings of our simulation study (that are given below), we will not include a comparison with the HD estimator, since it is not an aim of our paper to investigate
for these other settings whether the definition of randomized p-value of \cite{Dickhaus:2013,Hoang:2022a} leads to valid randomized p-values that can be practically computed.

The proposed
estimators will use the triangular density $\omega$ (mentioned in
\autoref{ThmAllConstructions}) to reduce a bit of the computations in their
implementations. Further, the double integral in the construction of the
proposed estimators is computed by an iterated integral for which each single
integral is approximated by a Riemann sum based on an equally spaced partition
with norm $0.01$, so as to reduce a bit the computational complexity of the
estimators when the number of hypotheses to test is very large. However, we
will not explore here how much more accurate these estimators can be when
finer partitions are used to obtain the Riemman sums.

For an estimator $\hat{\pi}_{1,m}$ of $\pi_{1,m}$, its accuracy is measured by
the excess $\tilde{\delta}_{m}=\hat{\pi}_{1,m}\pi_{1,m}^{-1}-1$. For each
experiment, the mean $\mu_{m}^{\ast}$ and standard deviation $\sigma_{m}%
^{\ast}$ of $\tilde{\delta}_{m}$ is estimated from independent realizations of
the experiment. Among two estimators of $\pi_{1,m}$, the one that has smaller
$\sigma_{m}^{\ast}$ is taken to be more stable, and the one that has both
smaller $\sigma_{m}^{\ast}$ and smaller $\left\vert \mu_{m}^{\ast}\right\vert
$ is better. On the other hand, in terms of non-asymptotic conservative FDR
control for an adaptive single-step MTP, an estimator $\hat{\pi}_{1,m}$ of
$\pi_{1,m}$ is preferred to under-estimate $\pi_{1,m}$, i.e., it is preferred
that $1-\hat{\pi}_{1,m}$ over-estimates $\pi_{0,m}$. This fact can be found,
e.g., in Section 3.2 of \cite{Blanchard:2009}. Therefore, among two estimators
of $\pi_{1,m}$ for a fixed $m$, the one that has smaller $\sigma_{m}^{\ast}$
and larger, non-positive $\mu_{m}^{\ast}$ will be considered better for
non-asymptotic conservative FDR control.

In the remainder of this section, \autoref{simDesign} contains a simulation
study for the setting of a point null, \autoref{simDesignDep} a simulation
study for the settings of bounded and one-sided nulls, and \autoref{SecSimRob}
a simulation study on the robustness of the proposed proportion estimators for
each of the three types of nulls.

\subsection{Simulations for the case of a point null}

\label{simDesign}

For the setting of a point null, we will examine the performances of the
estimator $\hat{\varphi}_{m}\left(  t_{\gamma,m},\mathbf{z}\right)  $ (denoted
by $\hat{\pi}_{\mathrm{J}}$) with $t_{\gamma,m}=\sqrt{2\sigma_{\left(
m\right)  }^{-1}\gamma\ln m}$ and $\sigma_{\left(  m\right)  }\equiv1$ and the
adaptive single-step MTP given in \autoref{ConstFDREst}. Set $\mathbf{z}%
\sim\mathcal{N}_{m}\left(  \boldsymbol{\mu},\boldsymbol{\Sigma}\right)  $
where $\boldsymbol{\Sigma}$ is a correlation matrix $\mathbf{R}=\left(  {\rho
}_{ij}\right)  $. We consider $10$ values for $m$ as $10^{3}$, $2\times10^{3}%
$, $4\times10^{3}$, $6\times10^{3}$, $8\times10^{3}$, $10^{4}$, $2\times
10^{4}$, $4\times10^{4}$, $8\times10^{4}$ and $10^{5}$. Recall $m_{0}$ as the
number of zero $\mu_{i}$'s and $\pi_{1,m}=1-m_{0}m^{-1}$ as the proportion of
nonzero $\mu_{i}$'s. Following the classification of sparsity by
\cite{Jin:2008}, we consider $4$ sparsity levels for $\pi_{1,m}$, i.e., the
dense regime $\pi_{1,m}=0.05$, moderately sparse regime $\pi_{1,m}=m^{-0.2}$,
critically sparse regime $\pi_{1,m}=m^{-0.5}$ and very sparse regime
$\pi_{1,m}=m^{-0.7}$. The nonzero $\mu_{i}$'s are generated independently such
that their absolute values $|\mu_{i}|$ are from the uniform distribution on
the compact interval $\left[  0.5,3.5\right]  $ but each $\mu_{i}$ has
probability $0.5$ to be negative or positive.

There are $3$ types of dependence, given below:

\begin{itemize}
\item \textquotedblleft Autoregressive\textquotedblright: ${\rho}_{ij}%
=\rho^{\left\vert i-j\right\vert }1_{\left\{  i\neq j\right\}  }$ with
$\rho=0.7$, i.e., $\mathbf{z}$ inherits the autocorrelation structure of an
autoregressive model of order $1$, such that $m^{-2}\left\Vert \mathbf{R}%
\right\Vert _{1}=\frac{1+\rho}{1-\rho}m^{-1}+O\left(  m^{-2}\right)  $ for
$\rho\in\left(  0,1\right)  $.

\item \textquotedblleft Long Range\textquotedblright: ${\rho}_{ij}=\rho=0.7$
for $i=1$ and $j=m-\left[  \sqrt{m}\right]  +1,...,m$ and ${\rho}_{ij}=0$ for
all other distinct pairs $\left(  i,j\right)  $. Namely, $z_{1}$ is equally
correlated with each $z_{j}$ for $j=m-\left[  \sqrt{m}\right]  +1,...,m$,
inducing long range dependence, and $m^{-2}\left\Vert \mathbf{R}\right\Vert
_{1}=m^{-1}+O\left(  m^{-3/2}\right)  .$

\item \textquotedblleft Moving Average\textquotedblright: $\mathbf{R}$ is a
banded matrix of bandwidth $\kappa=\left[  0.5\sqrt{m}\right]  $, such that
${\rho}_{ij}=\sum_{l=1}^{\kappa-\left\vert i-j\right\vert }\kappa_{l}%
\kappa_{\left\vert i-j\right\vert +l}$ with $\kappa_{i}=\kappa^{-1/2}$ when
$0<|i-j|<\kappa$. Namely, $\mathbf{z}$ inherits the autocorrelation structure
of a moving average model of order $\kappa$. The smallest off-diagonal nonzero
entry of $\mathbf{R}$ is $\kappa^{-1}$, and $m^{-2}\left\Vert \mathbf{R}%
\right\Vert _{1}=\frac{7}{24}m^{-1/2}+O\left(  m^{-1}\right)  $.
\end{itemize}

For the simulation design stated above, the Moving Average and Autoregressive
dependencies are respectively short range and strongly mixing, whereas the
Long Range dependence is neither strongly mixing nor short range. However, all
three types of dependence satisfy PCS and in the settings here are weak
dependence (with $\beta=\delta$ in (\ref{eq23}) and (\ref{24})) induced by PCS
and bounded maximal variance, and none of them has been considered in the
simulation studies of \cite{Meinshausen:2006}, \cite{Jin:2007} or
\cite{Jin:2008}. Further, neither of the critically sparse and very sparse
regimes was considered by \cite{Jin:2007} or \cite{Jin:2008}. Recall that
$\tilde{u}_{m}$ is the minimum of the absolute values of the nonzero Normal
means, and that the condition $\lim_{m\rightarrow\infty}\tilde{u}_{m}%
\sqrt{2\gamma\ln m}=\infty$ is sufficient to ensure the consistency of
$\hat{\pi}_{\mathrm{J}}$. Here $\tilde{u}_{m}=0.5$. So, the condition is satisfied.

There are a total of $120$ different experiments determined by the triple
$\left(  m,\pi_{1,m},\boldsymbol{\Sigma}\right)  $, and the simulation is
implemented by independently repeating $500$ times each experiment.

\subsubsection{Accuracy and stability of two proportion estimators}

\begin{figure}[th]
\centering
\includegraphics[height=0.25\textheight, width=1\textwidth]{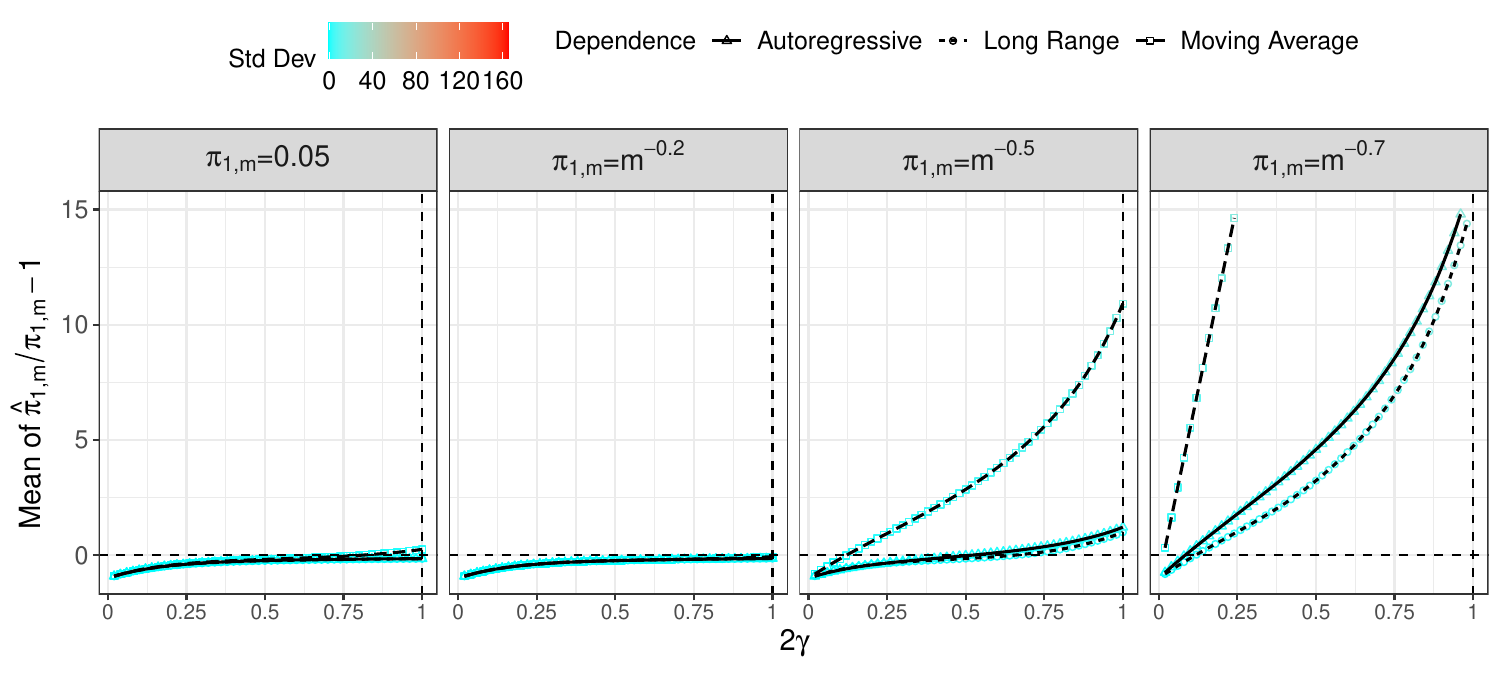}\caption[100000,
gamma varies, all]{Mean and standard deviation (``Std Dev'' in the color
legend) of the excess $\tilde{\delta}_{m}=\hat{\pi}_{1,m}\pi_{1,m}^{-1}-1$ of
Jin's estimator $\hat{\pi}_{\mathrm{J}}$ of \cite{Jin:2008} with thresholding
sequence $t_{\gamma,m}=\sqrt{2 \gamma\ln{m}}$ for $m=10^{5}$ as $2 \gamma$
changes. The curve for ``Moving Average'' dependence and $\pi_{1,m}=m^{-0.7}$
has been truncated at $15$ for the mean of $\tilde{\delta}_{m}$ to accommodate
the scales of other subfigures since for this curve the mean of $\tilde
{\delta}_{m}$ displays a monotone increasing relationship with $2 \gamma$ and
reaches around $112$ with its variance around $165$ when $2\gamma=1$.}%
\label{figGamma}%
\end{figure}

First, we discuss how Jin's estimator $\hat{\pi}_{\mathrm{J}}$ with
thresholding sequence $t_{\gamma,m}=\sqrt{2\gamma\ln m}$ behaves as $\gamma$
changes. \autoref{figGamma} shows the effects of the tuning parameter $\gamma$
on $\hat{\pi}_{\mathrm{J}}$ when $m=10^{5}$ and $\gamma$ ranges from $0.01$ to
$0.5$ with stepsize $0.01$. Two observations can be made: (1) When $\hat{\pi
}_{\mathrm{J}}$ is not applied to the very sparse regime or Moving Average
dependence and $\gamma$ changes within the theoretical range $\left(
0,2^{-1}\beta\right)  $ where $\beta$ is the index of weak dependence,
$\hat{\pi}_{\mathrm{J}}$ usually under-estimates $\pi_{1,m}$, the variance of
the excess $\tilde{\delta}_{m}$ usually has a smaller magnitude than its bias,
and $\hat{\pi}_{\mathrm{J}}$ becomes more accurate as $\gamma$ increases. (2)
In the critically sparse and very sparse regimes, if $\gamma$ is much larger
than $2^{-1}\beta$, $\hat{\pi}_{\mathrm{J}}$ can over-estimate $\pi_{1,m}$,
and the larger $\gamma$ is, the more it over-estimates $\pi_{1,m}$. However,
for some $\gamma$ in $\left(  0,2^{-1}\beta\right)  $, $\hat{\pi}_{\mathrm{J}%
}$ may still accurately estimate $\pi_{1,m}$ in the very sparse regime if it
is not applied to Moving Average dependence. This is likely due to the fact
that the bounds on $\left\vert e_{m}(t)\right\vert $ obtained through
$\left\Vert \boldsymbol{\Sigma}\right\Vert _{1}$ are not necessarily tight
when $\boldsymbol{\Sigma}$ is not diagonal.

\begin{figure}[th]
\centering
\includegraphics[height=0.5\textheight, width=1\textwidth]{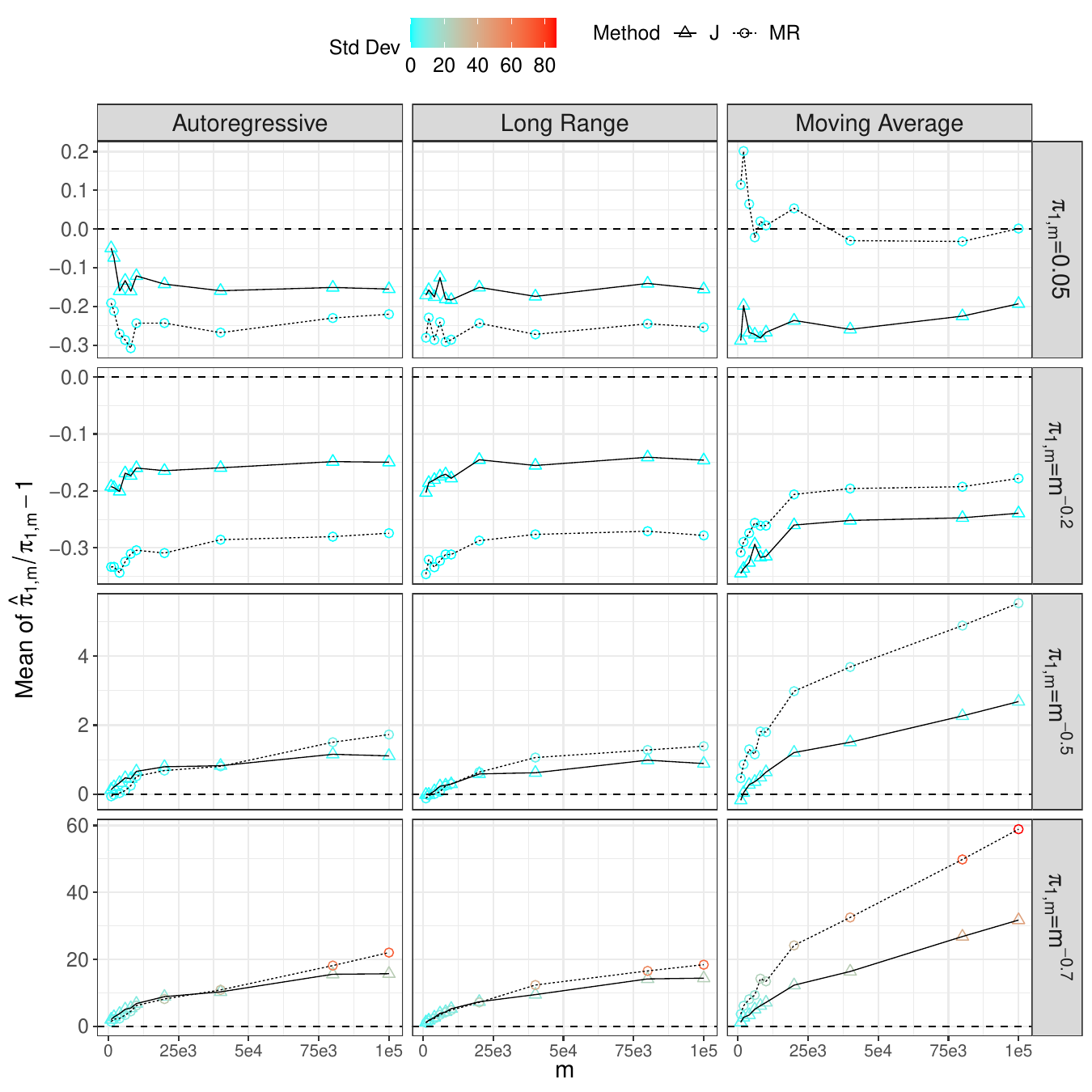}
\caption[unit variance, Jin]{Mean and standard deviation (``Std Dev'' in the
color legend) of the excess $\tilde{\delta}_{m}=\hat{\pi}_{1,m}\pi_{1,m}%
^{-1}-1$ of Jin's estimator (``J'') of \cite{Jin:2008} and the estimator
(``MR'') in \cite{Meinshausen:2006} when $m$ ranges from $10^{3}$ to $10^{5}%
$.}%
\label{figPiEstJin}%
\end{figure}

Second, we discuss the performances of $\hat{\pi}_{\mathrm{J}}$ and $\hat{\pi
}_{\mathrm{MR}}$. Specifically, $\hat{\pi}_{\mathrm{J}}$ is implemented with
$\gamma=0.24$ for Moving Average dependence with PCS index $\beta=0.5$ and
with $\gamma=0.49$ for Autoregressive and Long Range dependencies with PCS
index $\beta=1$. \autoref{figPiEstJin} shows the results. Note that the
independence case has been considered by \cite{Jin:2008} and
\cite{Meinshausen:2006}. For Autoregressive and Long Range dependencies and
all $4$ sparsity regimes, $\hat{\pi}_{\mathrm{J}}$ is accurate and is more
accurate than $\hat{\pi}_{\mathrm{MR}}$. In contrast, for Moving Average
dependence, $\hat{\pi}_{\mathrm{MR}}$ is more accurate than $\hat{\pi
}_{\mathrm{J}}$ in the dense and moderately sparse regimes, and is less
accurate than $\hat{\pi}_{\mathrm{J}}$ in the critically sparse and very
sparse regimes. In the critically sparse and very sparse regimes, both
$\hat{\pi}_{\mathrm{J}}$ and $\hat{\pi}_{\mathrm{MR}}$ can be unstable and
over-estimate $\pi_{1,m}$ when $m$ is large and $\beta$ is small. Further,
$\hat{\pi}_{\mathrm{MR}}$ may over-estimate $\pi_{1,m}$ in the dense regime
when $m$ is small.

\subsubsection{Accuracy of FDR estimation for adaptive multiple testing}

\begin{figure}[th]
\centering
\includegraphics[height=0.25\textheight, width=1\textwidth]{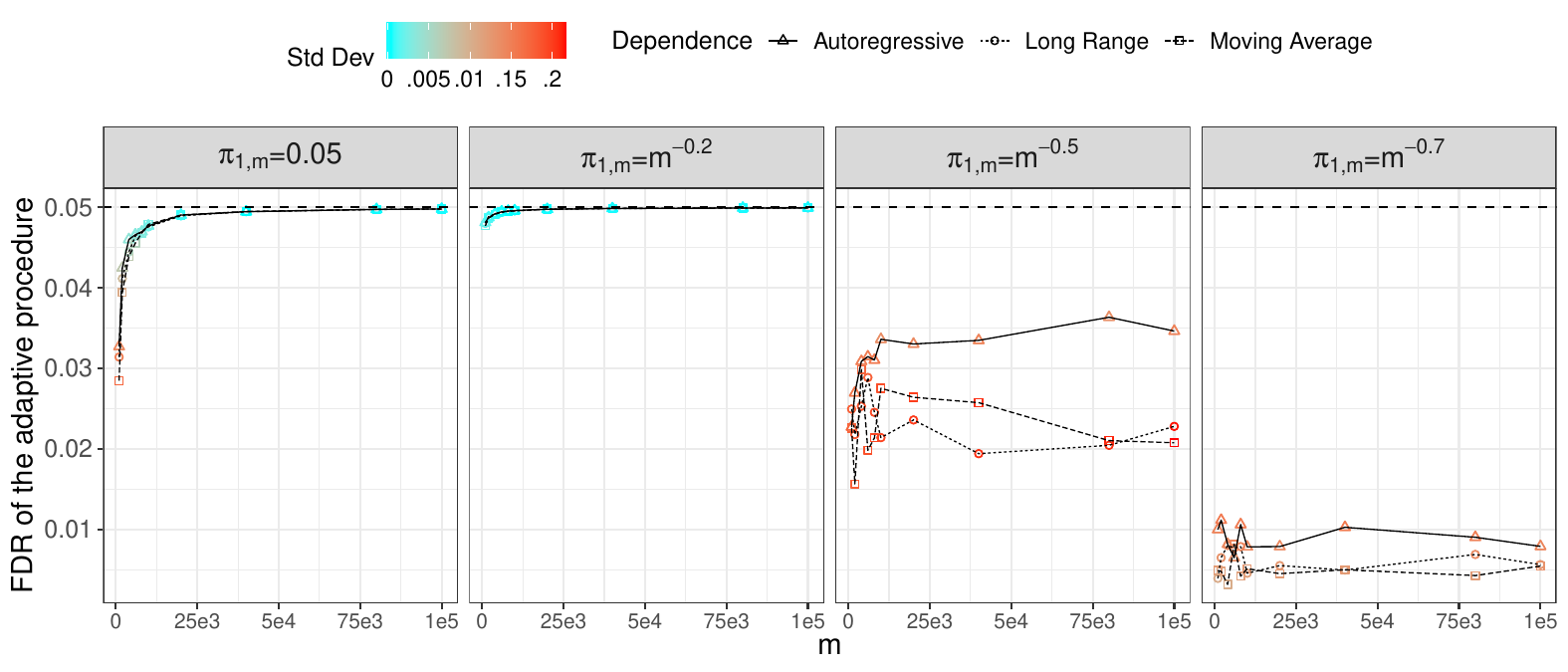}
\caption[fdr all]{The mean, i.e., FDR, and standard deviation of the FDP
(``Std Dev'' in the color legend) of the adaptive single-step MTP that employs
the proportion estimator $\hat{\pi}_{\mathrm{J}}$ of \cite{Jin:2008} as $m$
ranges from $10^{3}$ to $10^{5}$. Note that the horizontal line marks the
nominal FDR level $\alpha=0.05$.}%
\label{figFDPALL}%
\end{figure}

We discuss the accuracy of FDR estimation (based on
\autoref{ThmAdaptiveEstFDR}) for the adaptive single-step MTP introduced in
\autoref{ConstFDREst} that employs the estimator $\hat{\pi}_{\mathrm{J}}$ for
multiple testing the point null $\Theta_{0}=\left\{  0\right\}  $.
\autoref{figFDPALL} records the FDR of the adaptive MTP as $m$ increases,
where $\hat{\pi}_{\mathrm{J}}$ is implemented with $\gamma=0.24$ since the
smallest PCS index is $\beta=0.5$ for the $3$ types of dependence. The mean of
the FDP, i.e., the FDR, of the procedure is always upper bounded by the
nominal FDR level $\alpha=0.05$. It shows a strong trend of convergence to
$0.05$ as $m$ increases in the dense and moderately sparse regimes for all $3$
types of dependence, and it is smaller than $\alpha$ in the critically sparse
and very sparse regimes for all $3$ types of dependence even though the
estimator $\hat{\pi}_{\mathrm{J}}$ is not theoretically proven to be
consistent in these two sparsity regimes. This provides evidence that
asymptotic FDR control under weak dependence (that is induced by PCS) may be
robust to the accuracy of proportion estimation. However, in the critically
sparse and very sparse regimes, the FDP of the procedure is more variable,
likely due to the increased variability of the employed estimator $\hat{\pi
}_{\mathrm{J}}$.

\subsection{Simulations for the cases of one-sided and bounded nulls}

\label{simDesignDep}

We will check the performances of the proposed estimators, denoted now by
\textquotedblleft NEW\textquotedblright, by also comparing them with the
\textquotedblleft MR\textquotedblright\ estimator $\hat{\pi}_{\mathrm{MR}}$
for the settings of one-sided and point nulls. For the one-sided null
$\Theta_{0}=\left(  -\infty,0\right)  $, when $X_{0}$ is an observation from a
random variable $X$ with CDF $F_{\mu}$, $\mu\in U$, its one-sided p-value is
computed as $1-F_{0}\left(  X_{0}\right)  $.

We will simulate $\mathbf{z}\sim\mathcal{N}_{m}\left(  \boldsymbol{\mu
},\boldsymbol{\Sigma}\right)  $ with $\boldsymbol{\Sigma}$ being a correlation
matrix, where $\boldsymbol{\Sigma}$ represents a dependence structure given in
\autoref{simDesign}. We consider $6$ values for $m=10^{3}$, $5\times10^{3}$,
$10^{4}$, $5\times10^{4}$, $10^{5}$ or $5\times10^{5}$, and $2$ sparsity
levels $\pi_{1,m}=0.2$ (indicating the dense regime) or $\left(  \ln\ln
m\right)  ^{-1}$ (indicating the moderately sparse regime), where we recall
$\pi_{1,m}=m_{1}m^{-1}$ and $m_{0}+m_{1}=m$. The speed of the proposed
estimators $\hat{\varphi}_{m}\left(  t_{m},\mathbf{z}\right)  $ is set to be
$t_{m}=\sqrt{0.45\ln m}$ since the minimal of the weak dependence index
$\delta$ for the $3$ dependence structures is $0.5$, and $u_{m}=\tilde{u}%
_{m}=\left(  \ln\ln m\right)  ^{-1}$, where $u_{m}$ and $\tilde{u}_{m}$ are
respectively defined by \autoref{ThmIVaDep} and \autoref{ThmTypeI-VDep}. This
ensures $t_{m}^{-1}\left(  1+\max\left\{  u_{m}^{-1},\tilde{u}_{m}%
^{-1}\right\}  \right)  =o\left(  \pi_{1,m}\right)  $ and the consistency of
the proposed proportion estimators.

For $a<b$, let $\mathsf{U}\left(  a,b\right)  $ be the uniform random variable
or the uniform distribution on the closed interval $\left[  a,b\right]  $. For
the bounded null $\Theta_{0}=\left(  a,b\right)  $ or one-sided null
$\Theta_{0}=\left(  -\infty,0\right)  $, the $\mu_{i}$'s are generated as follows:

\begin{itemize}
\item Scenario 1 \textquotedblleft estimating $\pi_{1,m}$ for a bounded
null\textquotedblright: set $a=-1$ and $b=2$; generate $m_{0}$ $\mu_{i}$'s
independently from $\mathsf{U}\left(  a+u_{m},b-u_{m}\right)  $, $m_{11}$
$\mu_{i}$'s independently from $\mathsf{U}\left(  b+u_{m},b+6\right)  $, and
$m_{11}$ $\mu_{i}$'s independently from $\mathsf{U}\left(  a-4,a-u_{m}\right)
$, where $m_{11}=\max\left\{  1,\left[  0.5m_{1}\right]  -\left[  m/\ln\ln
m\right]  \right\}  $; set half of the remaining $m-m_{0}-2m_{11}$ $\mu_{i}$'s
to be $a$, and the rest to be $b$.

\item Scenario 2 \textquotedblleft estimating $\pi_{1,m}$ for a one-sided
null\textquotedblright: set $b=0$; generate $m_{0}$ $\mu_{i}$'s independently
from $\mathsf{U}\left(  -4,b-u_{m}\right)  $, and $\left[  0.9m_{1}\right]  $
$\mu_{i}$'s independently from $\mathsf{U}\left(  b+u_{m},b+6\right)  $; set
the rest $\mu_{i}$'s to be $b$.
\end{itemize}

\begin{figure}[th]
\centering
\includegraphics[height=0.8\textheight,width=1\textwidth]{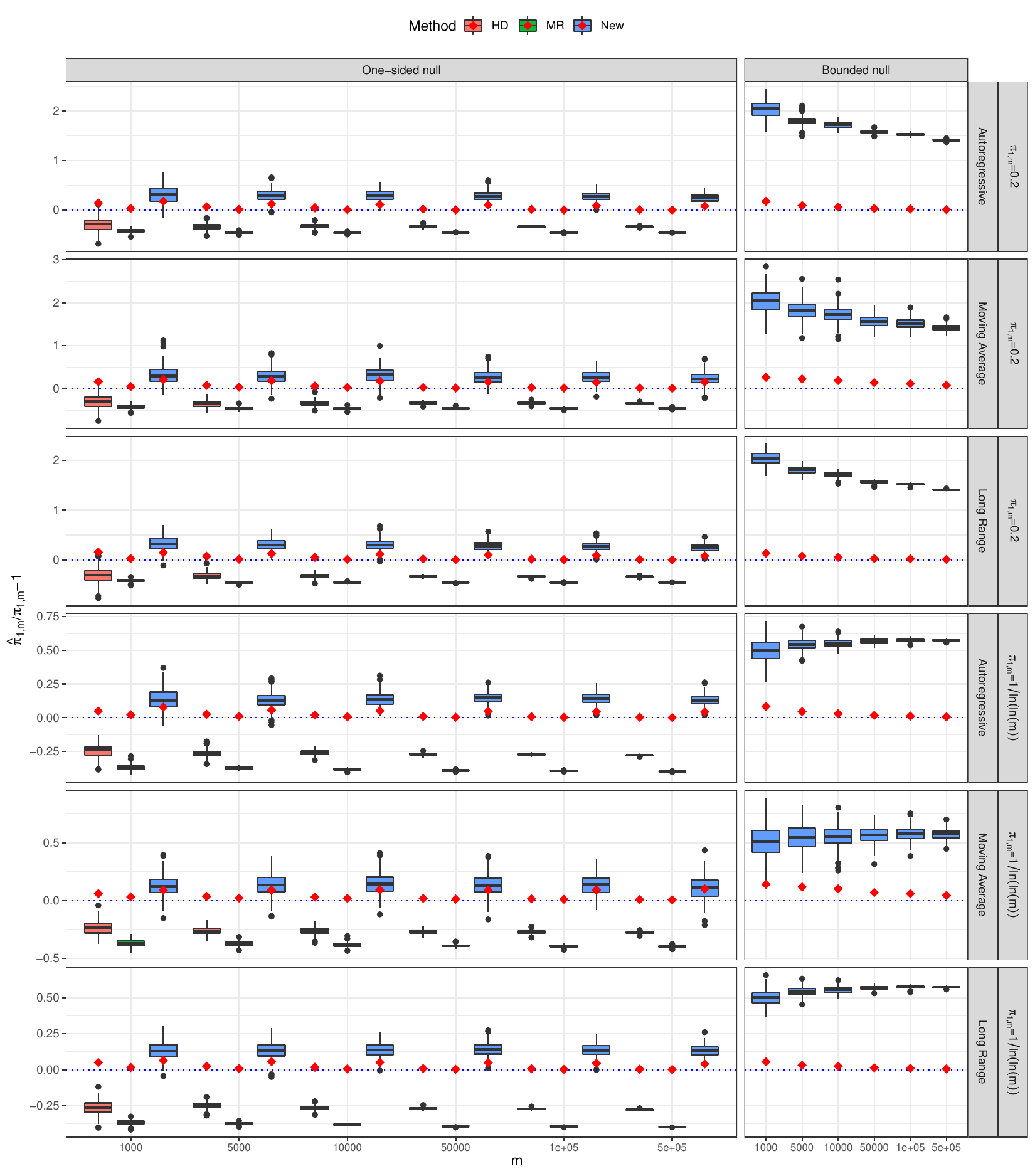}
\vspace{-0.3cm}\caption[estDep]{Boxplot of the excess $\tilde{\delta}_{m}%
=\hat{\pi}_{1,m}\pi_{1,m}^{-1}-1$ (on the vertical axis) of an estimator
$\hat{\pi}_{1,m}$ of ${\pi}_{1,m}$. The thick horizontal line and the diamond
in each boxplot are respectively the mean and standard deviation of
$\tilde{\delta}_{m}$, and the dotted horizontal line is the reference for
$\tilde{\delta}_{m}=0$. For for each $m$ in each setting for the one-side null, there
are three boxplots, where the leftmost is for the ``HD" estimator, the middle for the ``MR" estimator, and the rightmost for the proposed estimator ``New".
No simulation was done for the ``HD'' or ``MR'' estimator for the bounded null.}%
\label{figEstDep}%
\end{figure}

Scenario 1 and Scenario 2 are practical settings used in the simulations of
\cite{Chen:2019a}. There are a total of $72$ experiments, and each experiment
is independently repeated $250$ times. \autoref{figEstDep} visualizes the
performances of the proposed estimators and the MR estimator. For the point
and one-sided nulls respectively, the proposed estimators display strong
trends towards consistency, are stable and are more accurate than the MR
estimator, while for the bounded null, the proposed estimator displays a
strong trend of convergence and is stable and reasonably accurate. Note that
the MR estimator is often $0$ for the case of the one-sided null, failing to
detect the existence of alternative hypothesis. For one-side null, our proposed estimator performs
better than both the HD and MR estimators, and the HD estimator does better than the MR estimator.

\subsection{Simulations for Student's t random variables}

\label{SecSimRob}

We will consider all $3$ types of nulls in this subsection. Specifically, we
will take the same set of values for $m$ and $\pi_{1,m}$ and the same settings
for $\boldsymbol{\Sigma}=\left(  \sigma_{ij}\right) $ as given by
\autoref{simDesignDep}, together with one additional setting where
$\boldsymbol{\Sigma}$ is the identity matrix for ``Independent'' $\left\{
z_{i}\right\} _{i=1}^{m}$. As for the $\mu_{i}$'s, they are generated the same
way for the point null as in \autoref{simDesign} but the same way for the
bounded and one-sided nulls as in \autoref{simDesignDep}. The Student's t
random variables are generated as follows:\ set $n=50$ and independently
generate $n$ $\mathbf{z}_{j}\sim\mathcal{N}_{m}\left(  \boldsymbol{\mu
},\boldsymbol{\Sigma}\right)  $. For each $j$ let $\mathbf{z}_{j}=\left(
z_{j1},\ldots,z_{jm}\right)  ^{\top}$, and for each $i$ let $\xi_{i}=\bar
{z}_{i}/\left(  s_{i}/\sqrt{n}\right)  $, where $\bar{z}_{i}=n^{-1}\sum
_{j=1}^{n}z_{ji}$ and $s_{i}=\left(  n-1\right)  ^{-1}\sum_{j=1}^{n}\left(
z_{ji}-\bar{z}_{i}\right)  ^{2}$. So, each $\xi_{i}$ follows a Student's t
distribution but will be approximately Normal (since $n=50$). We refer to
$\left\{  \xi_{i}\right\}  _{i=1}^{m}$ as \textquotedblleft weakly
dependent\textquotedblright\ since they are induced by weakly dependent Normal
random variables. To assess the robustness of the proposed estimators, we
apply them to $\left\{  \xi_{i}\right\}  _{i=1}^{m}$. The speeds of
convergence of the proposed estimators are set to be $t_{m}=\sqrt{0.45\ln m}$
since the minimal of the weak dependence index $\delta$ for the three
dependence structures encoded by $\boldsymbol{\Sigma}$ is $0.5$. There are a
total of $144$ experiments, and each experiment is independently repeated
$250$ times. For the point and one-sided nulls respectively, the p-values to
which the MR estimator is applied are computed using the central Student t
distribution associated with the $\xi_{i}$'s (as usual for the point null but
as given by \autoref{simDesignDep} for the one-sided null, respectively).

Let $\lambda_{i}$ be the mean of $\xi_{i}$ for each $i$. Since $\sigma_{ii}%
=1$,%
\[
\xi_{i}=\frac{\bar{z}_{i}-\mu_{i}}{s_{i}/\sqrt{n}}+\frac{\mu_{i}}{s_{i}%
/\sqrt{n}},
\]
and $\left(  \bar{z}_{i}-\mu_{i}\right)  /\left(  s_{i}/\sqrt{n}\right)  $ has
mean $0$, we have%
\begin{equation}
\lambda_{i}=\mu_{i}\sqrt{n\left(  n-1\right)  }\mathbb{E}\left\{  \chi
_{n-1}^{2\times\left(  -1/2\right)  }\right\}  =\mu_{i}\sqrt{n\left(
n-1\right)  }\frac{\Gamma\left(  2^{-1}n-1\right)  }{\sqrt{2}\Gamma\left(
2^{-1}n-2^{-1}\right)  }, \label{eq11cc}%
\end{equation}
where $\chi_{n-1}^{2}$ is the central Chi-squared random variable with $n-1$
degrees of freedom. So, for the point null $\Theta_{0}=\left\{  0\right\}  $
and one-sided null $\Theta_{0}=\left\{  \mu:\mu<0\right\}  $ we must have%
\[
\pi_{1,m}=1-m^{-1}\left\vert \left\{  i\in\left\{  1,\ldots,m\right\}
:\lambda_{i}\in\Theta_{0}\right\}  \right\vert ,
\]
i.e., the alternative proportion associated with $\left\{  \mu_{i}\right\}
_{i=1}^{m}$ for the generating Normal random vector $\mathbf{z}\sim
\mathcal{N}_{m}\left(  \boldsymbol{\mu},\boldsymbol{\Sigma}\right)  $ is the
same as that for the means $\left\{  \lambda_{i}\right\}  _{i=1}^{m}$ of the
induced Student's t random variables $\left\{  \xi_{i}\right\}  _{i=1}^{m}$,
and we are accessing the robustness of the proportion estimators against
violations of distributional assumptions. In contrast, for the bounded null
$\Theta_{0}=\left\{  \mu:\mu\in\left(  a,b\right)  \right\}  $, the
alternative proportion $\pi_{1,m}^{\ast\ast}$ associated with the means
$\left\{  \lambda_{i}\right\}  _{i=1}^{m}$ is%
\[
\pi_{1,m}^{\ast\ast}=1-m^{-1}\left\vert \left\{  i\in\left\{  1,\ldots
,m\right\}  :\lambda_{i}\in\left(  a,b\right)  \right\}  \right\vert ,
\]
which in view of (\ref{eq11cc}) is not equal to the alternative proportion
$\pi_{1,m}$ associated with $\left\{  \mu_{i}\right\}  _{i=1}^{m}$.
Specifically, when $n=50$ and $\left(  a,b\right)  =\left(  -1,2\right)  $, we
have $\lambda_{i}\approx$ $7.181\mu_{i}$ and that $\pi_{1,m}^{\ast\ast}$ is
much larger than $\pi_{1,m}$ for each fixed $m$ and is very close to $1$. So,
for the bounded null, we are accessing the robustness of the proposed
estimators against violations of distributional assumptions and misspecified
alternative proportions.

\begin{figure}[th]
\centering
\includegraphics[height=0.85\textheight,width=1\textwidth]{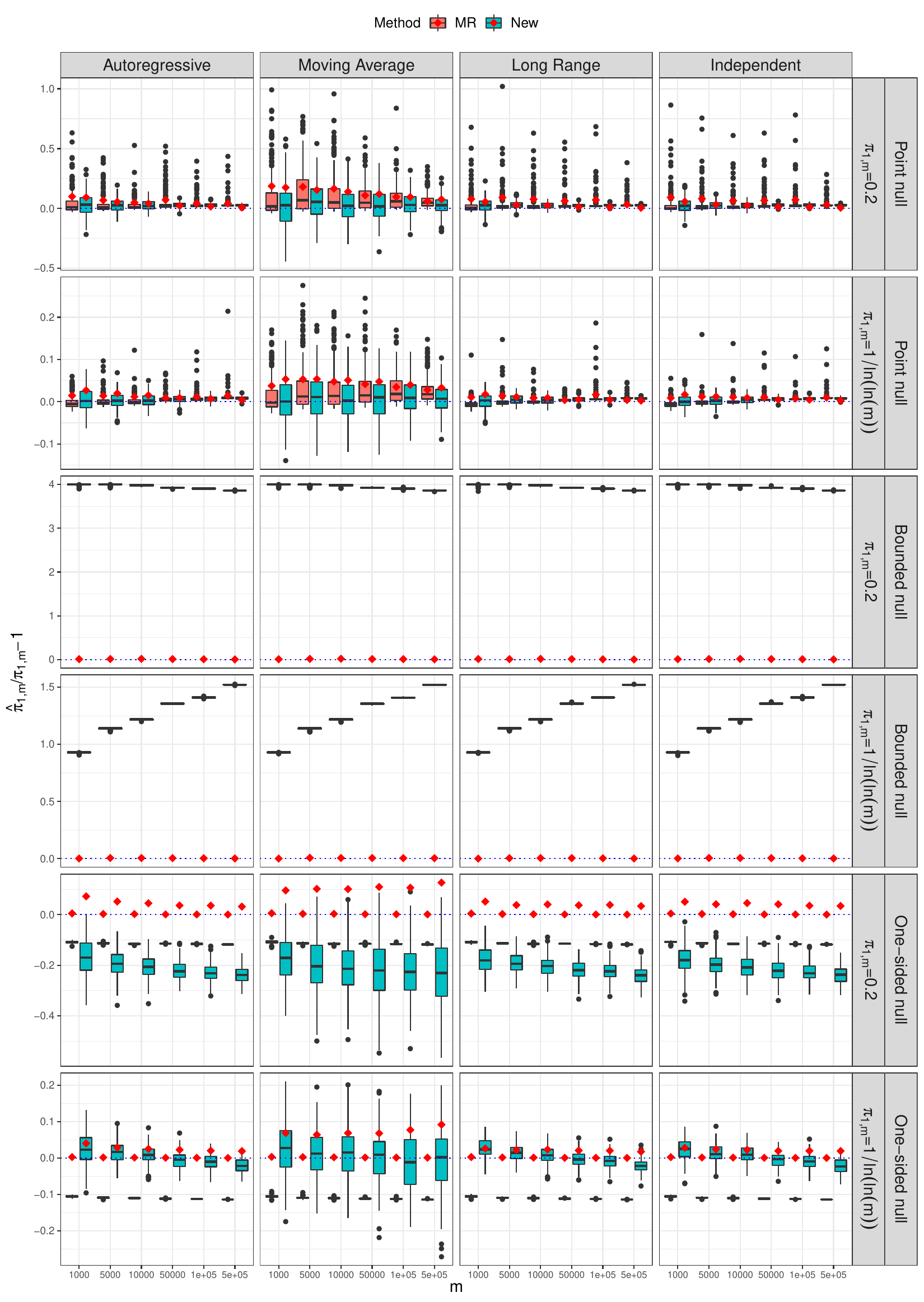}
\vspace{-0.4cm}\caption[Rob point]{Boxplot of the excess $\tilde{\delta}%
_{m}=\hat{\pi}_{1,m}\pi_{1,m}^{-1}-1$ (on the vertical axis) of an estimator
$\hat{\pi}_{1,m}$ of ${\pi}_{1,m}$. The thick horizontal line and the diamond
in each boxplot are respectively the mean and standard deviation of
$\tilde{\delta}_{m}$, and the dotted horizontal line is the reference for
$\tilde{\delta}_{m}=0$. For each pair of boxplots for each $m$ for the point
null and one-sided null respectively, the right one is for the proposed
estimator ``New'' and the left one the ``MR'' estimator. No simulation on the
``MR'' estimator was done for the bounded null.}%
\label{figEstRobAll}%
\end{figure}

\autoref{figEstRobAll} visualizes the results for the settings of the point,
one-sided and bounded nulls. The following can be observed: (i) for the point
null, both the proposed estimator and MR estimator are accurate and stable,
with the former being more accurate and more stable than the latter. (ii) for
the one-sided null, the MR estimator is more accurate than the proposed
estimator in the dense regime, whereas the opposite holds in the moderately
sparse regime. Specifically, the new estimator seems to be very accurate in
the moderately sparse regime, even though it may under-estimate the
alternative proportion when $m$ is very large. (iii) for the bounded null, as
explained earlier, the actual alternative proportion $\pi_{1,m}^{\ast\ast}$ is
much larger than $\pi_{1,m}$ for each fixed $m$ and is very close to $1$ in
both the dense and moderately sparse regimes. In view of this, the proposed
estimator is reasonably accurate in terms of estimating $\pi_{1,m}^{\ast\ast}$
(rather than $\pi_{1,m}$) if we convert $\hat{\pi}_{1,m}\pi_{1,m}^{-1}-1$ into
$\hat{\pi}_{1,m}\left(  \pi_{1,m}^{\ast\ast}\right)  ^{-1}-1$ in
\autoref{figEstRobAll}. However, for the setting of a bounded null in the
moderately sparse regime where $\pi_{1,m}=\left(  \ln\ln m\right)  ^{-1}$, the
proposed estimator does not show a strong trend towards consistency due to
violations of assumptions and a misspecified alternative proportion. Note that
the accuracy of the proportion estimators can be improved by using more
refined Riemman sums as mentioned at the beginning of
\autoref{SecNumericalStudies}.

\clearpage

\section{Discussion}

\label{SecConcAndDisc}

Due to issues related to defining and computing a p-value and the difficulty
in estimating the expectation of the average number of false rejections when
testing a composite null, we were not able to construct an adaptive
single-step MTP for testing composite nulls in general whose FDP satisfies the
weak law of large numbers and hence consistently estimates its FDR under weak
dependence or PCS. Even though multiple testing composite nulls is more
relevant in practice than multiple testing a point null, theoretical
progresses in FDR\ control and estimation under dependence for adaptive MTPs
for the former seems to be insufficient. We hope our assertions on adaptive
single-step MTPs under a PCS and consistency results on the proportion
estimators under weak dependence can be useful for further developments in
adaptive multiple testing composite nulls and under dependence.

For the proportion estimators that have been discussed here, we have focused
on their speeds of convergence at the maximal order $\sqrt{\ln m}$ and their
uniform consistency at these speeds. In practice, this allows these estimators
to achieve faster convergence to the proportion of false nulls than using a
slower speed of convergence. However, when these estimators employ slower
speeds of convergence, they will be able to consistently estimate sparser
proportions of false nulls than when they use faster speeds of convergence.
This principle applies to all proportion estimators of \cite{Jin:2007},
\cite{Jin:2008}, \cite{Chen:2018a} and \cite{Chen:2019a}. On the other hand,
these proportion estimators can be used to estimate the sparsity level of the
vector of regression parameters in high-dimensional sparse linear models with
weakly dependent Gaussian errors, similar to the setting discussed by
\cite{Chen:2019a} for such models with independent Gaussian random errors.

\appendix

\section{Proofs}

\label{secProofs}

We provide the proofs of \autoref{prop:SLLNComp}, \autoref{ThmAdaptiveEstFDR},
\autoref{LmBoundDiffPhaseFuncWTumix}, \autoref{ThmConsistPlugEst},
\autoref{ThmIVaDep} and \autoref{ThmTypeI-VDep}.

\subsection{Proof of \autoref{prop:SLLNComp}}

First, consider the case $\Theta_{0}=\left\{  0\right\}  $. Since
$\mathrm{cov}\big(1_{\left\{  p_{i}\leq\tau\right\}  },1_{\left\{  p_{j}%
\leq\tau\right\}  }\big)$ for two right-tailed p-values $p_{i}$ and $p_{j}$ is
the same as that for two left-tailed p-values $p_{i}$ and $p_{j}$, the
inequality (\ref{IneqA}), i.e.,%
\[
\left\vert \mathrm{cov}\big(1_{\left\{  p_{i}\leq\tau\right\}  },1_{\left\{
p_{j}\leq\tau\right\}  }\big)\right\vert \leq C\left\vert \rho_{ij}\right\vert
\text{ for all }i\neq j
\]
holds for left-tailed p-values $\left\{  p_{i}\right\}  _{i=1}^{m}$ that are
associated with $\left\{  z_{i}\right\}  _{i=1}^{m}$.

Second, consider the case where $\Theta_{0}$ is not a singleton and%
\[
p_{i}=\sup\nolimits_{\mu\in\Theta_{0}}F_{\mu}\left(  z_{i}\right)  \text{
where }F_{\mu}\text{ is the CDF of }X\sim\mathcal{N}_{1}\left(  \mu
,\sigma_{ii}\right)
\]
for each $i$. Then, for each p-value $p_{i}=\sup_{\mu\in\Theta_{0}}F_{\mu
}\left(  z_{i}\right)  $, we can pick a sequence $\mu_{i,l}\in\Theta_{0}$ and
set $p_{i,l}=F_{\mu_{i,l}}\left(  z_{i}\right)  $ such that $p_{i}%
=\lim_{l\rightarrow\infty}p_{i,l}$. However, we see from the proof of
\autoref{prop:SLLNMarginal} by \cite{Chen:2014SLLN} that, for each pair
$\left(  z_{i},z_{j}\right)  $ with $i\neq j$, (\ref{IneqA}) also holds
uniformly in $\left(  \mu_{i},\mu_{j}\right)  \in\mathbb{R}^{2}$ and $\left(
\sigma_{ii},\sigma_{jj}\right)  \in\mathbb{R}^{2}$. Therefore,%
\[
\left\vert \mathrm{cov}\big(1_{\left\{  p_{i,l}\leq\tau\right\}  },1_{\left\{
p_{i,l}\leq\tau\right\}  }\big)\right\vert \leq C\left\vert \rho
_{ij}\right\vert \text{ for all }i\neq j,
\]
and applying the dominated convergence theorem (DCT), we obtain%
\[
\mathrm{cov}\big(1_{\left\{  p_{i}\leq\tau\right\}  },1_{\left\{  p_{j}%
\leq\tau\right\}  }\big)=\lim_{l\rightarrow\infty}\mathrm{cov}\big(1_{\left\{
p_{i,l}\leq\tau\right\}  },1_{\left\{  p_{i,l}\leq\tau\right\}  }\big),
\]
and (\ref{IneqA}) holds for these p-values.

Third, the proof of \autoref{prop:SLLNMarginal} reveals the following: when
(\ref{IneqA}) and (\ref{SLLNAB}) hold, the condition $\liminf_{m\rightarrow
\infty}m^{-1}R_{m}\left(  t\right)  >0$ implies (\ref{SLLNC}), i.e.,
$\lim_{m\rightarrow\infty}\left\vert \mathrm{FDP}_{m}\left(  \tau\right)
-\mathbb{E}\left[  \mathrm{FDP}_{m}\left(  \tau\right)  \right]  \right\vert
=0$ almost surely, regardless of whether $\Theta_{0}$ is singleton or not.
However, if $\liminf_{m\rightarrow\infty}\pi_{0,m}>0$ and $\lim_{m\rightarrow
\infty}\min_{i\in I_{0,m}}\mathbb{P}\left(  p_{i}\leq\tau\right)  >0$, then%
\begin{align*}
\liminf_{m\rightarrow\infty}\frac{R_{m}\left(  \tau\right)  }{m}  &
\geq\liminf_{m\rightarrow\infty}\frac{m_{0}}{m}\frac{V_{m}\left(  \tau\right)
}{m_{0}}\geq\liminf_{m\rightarrow\infty}\pi_{0,m}\liminf_{m\rightarrow\infty
}\frac{V_{m}\left(  \tau\right)  }{m_{0}}\\
&  \geq C\frac{m_{0}\lim_{m\rightarrow\infty}\min_{i\in I_{0,m}}%
\mathbb{P}\left(  p_{i}\leq\tau\right)  }{2m_{0}}>0
\end{align*}
for some constant $C>0$. This completes the proof.

\subsection{Proof of \autoref{ThmAdaptiveEstFDR}}

Firstly, $\alpha_{m}\left(  \hat{\tau}\right)  \leq\alpha$ almost surely by
the definition of $\hat{\tau}$. Let $r_{m}=\min_{1\leq k\leq m}k^{-1}%
R_{k}\left(  \hat{\tau}\right)  $. Then $\mathbb{P}\left(  r_{m}>0\right)
\rightarrow1$. Since the processes $m_{0}^{-1}V_{m}\left(  t\right)  $ and
$m^{-1}R_{m}\left(  t\right)  $ and the FDP process are all nonnegative and
uniformly upper bounded by $1$ almost surely, we can assume $\mathbb{P}\left(
r_{m}>0\right)  =1$ without affecting any conclusion on convergence in
probability to be presented next. Let $\twoheadrightarrow$ denote
\textquotedblleft almost sure convergence\textquotedblright\ and
$\rightsquigarrow$ \textquotedblleft convergence in
probability\textquotedblright. We will frequently use
\autoref{prop:SLLNMarginal} and \autoref{prop:SLLNComp} but will not mention
this every time. In the rest of the proof, $m$ is sufficiently large and each
limit is for when $m\rightarrow\infty$.

Let $c\left(  t\right)  =\mathbb{P}\left(  p_{i_{0}}\leq\tau\right)  $ for
some $i_{0}\in I_{0,m}$. Since $\left\{  p_{i}\right\}  _{i\in I_{0,m}}$ are
identically distributed, then $m_{0}^{-1}V_{m}\left(  t\right)
\twoheadrightarrow c\left(  t\right)  $. When $m^{-1}\mathbb{E}\left[
R_{m}\left(  t\right)  \right]  \rightarrow Q\left(  t\right)  $ and
$\pi_{1,m}\rightarrow\pi_{1}$, we can define $\pi_{0}=1-\pi_{1}$,%
\[
A\left(  t\right)  =\frac{\pi_{0}c\left(  t\right)  }{Q\left(  t\right)
}\text{ \ \ and \ }\tau_{1}=\sup\left\{  t\in\left[  0,1\right]  :A\left(
t\right)  \leq\alpha\right\}  .
\]
Recall
\[
\hat{\tau}=\sup\left\{  t\in\left[  0,1\right]  :\alpha_{m}\left(  t\right)
=\frac{\left(  1-\hat{\pi}_{1,m}\right)  c\left(  t\right)  }{m^{-1}%
R_{m}\left(  t\right)  }\leq\alpha\right\}  .
\]
Then%
\[
\mathrm{FDP}_{m}\left(  \hat{\tau}\right)  =\frac{m^{-1}V_{m}\left(  \hat
{\tau}\right)  }{m^{-1}R_{m}\left(  \hat{\tau}\right)  }=\frac{\left(
1-\pi_{1,m}\right)  m_{0}^{-1}V_{m}\left(  \hat{\tau}\right)  }{m^{-1}%
R_{m}\left(  \hat{\tau}\right)  }%
\]
is the FDP of the adaptive procedure.

First, let us show $\hat{\tau}\rightsquigarrow\tau_{1}$. Since $\lim
_{t\downarrow0}A\left(  t\right)  \neq\alpha$, we have $\tau_{1}>0$. Further,
$A\left(  \tau_{1}+\varepsilon\right)  >\alpha$ must hold for any
$\varepsilon>0$ with $\tau_{1}+\varepsilon\in\left[  0,1\right]  $. Fix some
small $\delta_{1}>0$, and pick any $\delta_{2}\geq\delta_{1}$ such that
$t^{\prime}=\tau_{1}+\delta_{2}\in\left[  0,1\right]  $. Set%
\[
\Delta_{m}=\sup_{t\in\left[  0,1\right]  }\left\vert m^{-1}R_{m}\left(
t\right)  -Q\left(  t\right)  \right\vert \text{ \ and }\Gamma_{m}=\sup
_{t\in\left[  0,1\right]  }\left\vert m_{0}^{-1}V_{m}\left(  t\right)
-c\left(  t\right)  \right\vert .
\]
Since the almost sure limits $c\left(  t\right)  $ and $Q\left(  t\right)  $
are bounded and right continuous with left limits, by the same arguments in
the proof of the Glivenko-Cantelli theorem (see, e.g., page 20 of
\cite{Loeve:1977}), we have $\Delta_{m}\twoheadrightarrow0$ and $\Gamma
_{m}\twoheadrightarrow0$. However, $\left\vert \hat{\pi}_{1,m}-\pi
_{1,m}\right\vert \rightsquigarrow0$ and $\pi_{1,m}\rightarrow\pi_{1}$. So,%
\begin{align*}
\inf_{\delta_{2}\geq\delta_{1}}\frac{\left(  1-\hat{\pi}_{1,m}\right)
c\left(  t^{\prime}\right)  }{m^{-1}R_{m}\left(  t^{\prime}\right)  }  &
\geq\inf_{\delta_{2}\geq\delta_{1}}\frac{\left(  1-\hat{\pi}_{1,m}\right)
c\left(  t^{\prime}\right)  }{Q\left(  t^{\prime}\right)  +\Delta_{m}}\\
&  \geq\inf_{\delta_{2}\geq\delta_{1}}\frac{\pi_{0}c\left(  t^{\prime}\right)
-\varepsilon_{0}}{Q\left(  t^{\prime}\right)  +\Delta_{m}}>\alpha
\end{align*}
for some small $\varepsilon_{0}>0$. Thus, $\mathbb{P}\left(  \hat{\tau}%
<\tau_{1}+\delta_{1}\right)  \rightarrow1$ holds. On the other hand, the
derivative $\frac{d}{dt}A\left(  \tau_{1}\right)  $ must be positive when it
is defined, due to the definition of $\tau_{1}$. So, $A\left(  \tau_{1}%
-\delta_{3}\right)  <A\left(  \tau_{1}\right)  $ when $\delta_{3}>0$ is small
enough. Setting $t^{\prime\prime}=\tau_{1}-\delta_{4}$ by picking
$0<\delta_{4}\leq\delta_{3}$ yields%
\[
\frac{\left(  1-\hat{\pi}_{1,m}\right)  c\left(  t^{\prime\prime}\right)
}{m^{-1}R_{m}\left(  t^{\prime\prime}\right)  }\leq\frac{\left(  1-\hat{\pi
}_{1,m}\right)  c\left(  t^{\prime\prime}\right)  }{Q\left(  t^{^{\prime
\prime}}\right)  -\Delta_{m}}\leq\frac{\pi_{0}c\left(  t^{\prime\prime
}\right)  +\tilde{\varepsilon}_{0}}{Q\left(  t^{^{\prime\prime}}\right)
-\Delta_{m}}<\alpha
\]
for some small $\tilde{\varepsilon}_{0}>0$, again due to $\Delta
_{m}\twoheadrightarrow0$, $\left\vert \hat{\pi}_{1,m}-\pi_{1,m}\right\vert
\rightsquigarrow0$ and $\pi_{1,m}\rightarrow\pi_{1}$. Therefore,
$\mathbb{P}\left(  \hat{\tau}>\tau_{1}-\delta_{3}\right)  \rightarrow1$, and
$\hat{\tau}\rightsquigarrow\tau_{1}$ holds.

Second, let us show $\left\vert \alpha_{m}\left(  \hat{\tau}\right)  -A\left(
\tau_{1}\right)  \right\vert \rightsquigarrow0$. Clearly,%
\begin{align}
\left\vert m_{0}^{-1}V_{m}\left(  \hat{\tau}\right)  -c\left(  \tau
_{1}\right)  \right\vert  &  \leq\left\vert m_{0}^{-1}V_{m}\left(  \hat{\tau
}\right)  -c\left(  \hat{\tau}\right)  \right\vert +\left\vert c\left(
\hat{\tau}\right)  -c\left(  \tau_{1}\right)  \right\vert \nonumber\\
&  \leq\Gamma_{m}+\left\vert c\left(  \hat{\tau}\right)  -c\left(  \tau
_{1}\right)  \right\vert \rightsquigarrow0, \label{eqx5}%
\end{align}
and%
\begin{equation}
\left\vert m^{-1}R_{m}\left(  \hat{\tau}\right)  -Q\left(  \tau_{1}\right)
\right\vert \leq\Delta_{m}+\left\vert Q\left(  \hat{\tau}\right)  -Q\left(
\tau_{1}\right)  \right\vert \rightsquigarrow0, \label{eq31}%
\end{equation}
where the second term on the right-hand side (RHS) of each of (\ref{eqx5}) and
(\ref{eq31}) $\rightsquigarrow0$ due to $\hat{\tau}\rightsquigarrow\tau_{1}$
and the continuous mapping theorem (CMT). On the other hand, $\hat{\pi}%
_{1,m}\rightsquigarrow\pi_{1,m}$, $\pi_{1,m}\rightarrow\pi_{1}$ and
$\mathbb{P}\left(  r_{m}>0\right)  \rightarrow1$, which together with
(\ref{eqx5}), (\ref{eq31}) and the CMT imply%
\[
\alpha_{m}\left(  \hat{\tau}\right)  \rightsquigarrow A\left(  \tau
_{1}\right)  \text{ \ and \ }\mathrm{FDP}_{m}\left(  \hat{\tau}\right)
\rightsquigarrow A\left(  \tau_{1}\right)  .
\]
Thus, by the DCT, we must have $\left\vert \mathrm{FDP}_{m}\left(  \hat{\tau
}\right)  -\mathbb{E}\left[  \mathrm{FDP}_{m}\left(  \hat{\tau}\right)
\right]  \right\vert \rightarrow0$ and $\left\vert \alpha_{m}\left(  \hat
{\tau}\right)  -\mathbb{E}\left[  \alpha_{m}\left(  \hat{\tau}\right)
\right]  \right\vert \rightarrow0$, which also implies%
\[
\mathrm{FDP}_{m}\left(  \hat{\tau}\right)  \rightsquigarrow A\left(  \tau
_{1}\right)  \text{ \ and \ }\mathrm{FDR}_{m}\left(  \hat{\tau}\right)
\rightarrow A\left(  \tau_{1}\right)  .
\]
This completes the proof.

\subsection{Proof of \autoref{LmBoundDiffPhaseFuncWTumix}}

Recall $\sigma_{\left(  m\right)  }=\max_{1\leq i\leq m}\sigma_{ii}$ and
$\left\Vert \omega\right\Vert _{\infty}=\sup_{s\in\left[  -1,1\right]
}\left\vert \omega\left(  s\right)  \right\vert $. Let $S_{m}\left(  w\right)
=m^{-1}\sum\limits_{j=1}^{m}\cos\left(  wz_{j}\right)  $ and $d_{m}\left(
w\right)  =S_{m}\left(  w\right)  -\mathbb{E}\left[  S_{m}\left(  w\right)
\right]  $. Recall $e_{m}\left(  t\right)  =\hat{\varphi}_{m}\left(
t,\mathbf{z}\right)  -\varphi_{m}\left(  t,\boldsymbol{\mu}\right)  $. In
order to bound the oscillations of $\left\vert e_{m}\left(  t\right)
\right\vert $, we will bound the variance $\mathbb{V}\left[  d_{m}\left(
ts\right)  \right]  $ for $s\in\left[  -1,1\right]  $. Clearly, $\mathbb{V}%
\left[  d_{m}\left(  t\right)  \right]  =I_{1}+I_{2}$, where $I_{1}=m^{-2}%
\sum\limits_{i=1}^{m}\mathbb{V}\left[  \cos\left(  tsz_{i}\right)  \right]  $,
$I_{2}=2m^{-2}\sum\limits_{1\leq i<j\leq m}c_{ij}$ and $c_{ij}=\mathsf{cov}%
\left[  \cos\left(  tsz_{i}\right)  ,\cos\left(  tsz_{j}\right)  \right]  $
for $i\neq j$. Since $\left\vert \cos\left(  tsz_{i}\right)  \right\vert $ is
upper bounded by $1$ uniformly in $t,s$ and $z_{i}$, we see%
\begin{equation}
I_{1}=m^{-2}\sum\limits_{i=1}^{m}\mathbb{V}\left[  \cos\left(  tsz_{i}\right)
\right]  \leq m^{-1}. \label{eq7}%
\end{equation}
So, it is left to bound $I_{2}$. We divide the proof into two parts:
\textquotedblleft Part I\textquotedblright\ to upper bound $\left\vert
c_{ij}\right\vert $ for $i\neq j$; \textquotedblleft Part II\textquotedblright%
\ to upper bound the variance of $e_{m}\left(  t\right)  $.

Part I. Let $x_{i}=\exp\left(  \sqrt{-1}tsz_{i}\right)  $ for $1\leq i\leq m.$
Since $z_{i}+z_{j}$ is Normally distributed with mean $\mu_{i}+\mu_{j}$ and
variance $a_{ij}=\sigma_{ii}+\sigma_{jj}+2\sigma_{ij}$ for $i\neq j$, we have%
\[
\mathbb{E}\left[  x_{i}x_{j}\right]  =\exp\left(  \sqrt{-1}ts\left(
u_{i}+u_{j}\right)  \right)  \exp\left(  -2^{-1}t^{2}s^{2}a_{ij}\right)
\text{.}%
\]
Therefore, $\mathsf{cov}\left[  x_{i},x_{j}\right]  =\exp\left(  \sqrt
{-1}ts\left(  u_{i}+u_{j}\right)  \right)  g_{ij}$ with%
\begin{equation}
g_{ij}=\exp\left(  -2^{-1}t^{2}s^{2}\left(  \sigma_{ii}+\sigma_{jj}\right)
\right)  \left(  \exp\left(  -t^{2}s^{2}\sigma_{ij}\right)  -1\right)
\text{.} \label{eq9}%
\end{equation}
Similarly, since $z_{i}-z_{j}$ is Normally distributed with mean $\mu_{i}%
-\mu_{j}$ and variance $b_{ij}=\sigma_{ii}+\sigma_{jj}-2\sigma_{ij}$ for
$i\neq j$, we have $\mathsf{cov}\left[  x_{i},\overline{x_{j}}\right]
=\exp\left(  \sqrt{-1}ts\left(  u_{i}-u_{j}\right)  \right)  h_{ij}$, where
$\overline{x_{j}}$ is the complex conjugate of $x_{j}$ and%
\begin{equation}
h_{ij}=\exp\left(  -2^{-1}t^{2}s^{2}\left(  \sigma_{ii}+\sigma_{jj}\right)
\right)  \left(  \exp\left(  t^{2}s^{2}\sigma_{ij}\right)  -1\right)  \text{.}
\label{eq10}%
\end{equation}
Therefore,%
\begin{align}
c_{ij}  &  =\mathsf{cov}\left[  \cos\left(  tsz_{i}\right)  ,\cos\left(
tsz_{j}\right)  \right]  =2^{-1}\left[  \Re\left(  \mathsf{cov}\left[
x_{i},x_{j}\right]  \right)  +\Re\left(  \mathsf{cov}\left[  x_{i}%
,\overline{x_{j}}\right]  \right)  \right] \nonumber\\
&  =2^{-1}\left\{  g_{ij}\cos\left(  ts\left(  u_{i}+u_{j}\right)  \right)
+h_{ij}\cos\left(  ts\left(  u_{i}-u_{j}\right)  \right)  \right\}  ,
\label{eq8}%
\end{align}
where $\Re\left(  \cdot\right)  $ denotes the real part.

Since%
\begin{equation}
\left\vert \exp\left(  \pm t^{2}s^{2}\sigma_{ij}\right)  -1\right\vert
=t^{2}s^{2}\left\vert \sigma_{ij}\right\vert \exp\left(  t^{2}s^{2}\sigma
_{ij}^{\prime}\right)  \leq t^{2}s^{2}\left\vert \sigma_{ij}\right\vert
\exp\left(  t^{2}s^{2}\left\vert \sigma_{ij}\right\vert \right)  \label{eq11}%
\end{equation}
for some $\sigma_{ij}^{\prime}$ that lies strictly between $0$ and $\pm
\sigma_{ij}$, we see from (\ref{eq8}), (\ref{eq9}) and (\ref{eq10}) that, for
each fixed $t$ and $s$,%
\begin{align}
\left\vert c_{ij}\right\vert  &  \leq\max\left\{  g_{ij},h_{ij}\right\}
\nonumber\\
&  \leq t^{2}s^{2}\left\vert \sigma_{ij}\right\vert \exp\left(  -2^{-1}%
t^{2}s^{2}\left(  \sigma_{ii}+\sigma_{jj}\right)  \right)  \exp\left(
t^{2}s^{2}\left\vert \sigma_{ij}\right\vert \right) \nonumber\\
&  =t^{2}s^{2}\left\vert \sigma_{ij}\right\vert \exp\left(  -2^{-1}t^{2}%
s^{2}\left(  \sigma_{ii}+\sigma_{jj}-2\left\vert \sigma_{ij}\right\vert
\right)  \right)  . \label{eq12}%
\end{align}
By H\"{o}lder's inequality, $\left\vert \sigma_{ij}\right\vert \leq
\sqrt{\sigma_{ii}\sigma_{jj}}$. So,%
\[
\sigma_{ii}+\sigma_{jj}-2\left\vert \sigma_{ij}\right\vert \geq\sigma
_{ii}+\sigma_{jj}-2\sqrt{\sigma_{ii}\sigma_{jj}}=\left(  \sqrt{\sigma_{ii}%
}-\sqrt{\sigma_{jj}}\right)  ^{2}\geq0,
\]
and (\ref{eq12}) implies $\left\vert c_{ij}\right\vert \leq t^{2}%
s^{2}\left\vert \sigma_{ij}\right\vert $.

Part II. Therefore%
\[
\left\vert I_{2}\right\vert \leq2m^{-2}\sum\limits_{1\leq i<j\leq m}\left\vert
c_{ij}\right\vert \leq 2t^{2}s^{2}m^{-2}\sum_{1\leq i<j\leq m}\left\vert
\sigma_{ij}\right\vert \le 2t^{2}m^{-2}\left\Vert \mathbf{\Sigma}\right\Vert _{1}%
\]
and%
\begin{equation}
\mathbb{V}\left[  d_{m}\left(  ts\right)  \right]  \leq m^{-1}+2t^{2}%
m^{-2}\left\Vert \mathbf{\Sigma}\right\Vert _{1}. \label{eq17}%
\end{equation}

We are ready to bound $e_{m}\left(  t\right)  $. Set $p_{m}\left(  t\right)
=m^{-1}+2t^{2}m^{-2}\left\Vert \mathbf{\Sigma}\right\Vert _{1}$. By Markov's
inequality,%
\begin{equation}
\mathbb{P}\left(  \left\{  \left\vert d_{m}\left(  ts\right)  \right\vert
>\varepsilon\right\}  \right)  \leq\dfrac{\mathbb{V}\left[  d_{m}\left(
ts\right)  \right]  }{\varepsilon^{2}}\leq\dfrac{p_{m}\left(  t\right)
}{\varepsilon^{2}}\text{.} \label{eq19}%
\end{equation}
Recall $d_{m}\left(  w\right)  =S_{m}\left(  w\right)  -\mathbb{E}\left[
S_{m}\left(  w\right)  \right]  $. Since $\mathbb{E}\left[  \hat{\varphi}%
_{m}\left(  t,\mathbf{z}\right)  \right]  =\varphi_{m}\left(
t,\boldsymbol{\mu}\right)  $ for all $t$ and $\boldsymbol{\mu}$, we obtain%
\[
e_{m}\left(  t\right)  =m^{-1}\sum_{j=1}^{m}\int_{\left[  -1,1\right]  }%
\exp\left(  2^{-1}t^{2}s^{2}\sigma_{jj}\right)  \left(  \cos\left(
tsz_{j}\right)  -\mathbb{E}\left[  \cos\left(  tsz_{j}\right)  \right]
\right)  \omega\left(  s\right)  ds
\]
and%
\begin{equation}
\left\vert e_{m}\left(  t\right)  \right\vert \leq2\int_{\left[  0,1\right]
}\exp\left(  2^{-1}t^{2}s^{2}\sigma_{\left(  m\right)  }\right)  \left\vert
\omega\left(  s\right)  \right\vert \left\vert d_{m}\left(  ts\right)
\right\vert ds. \label{eqx1}%
\end{equation}
So, the RHS of (\ref{eqx1})\ and inequality (\ref{eq19}) imply
\begin{equation}
\left\vert e_{m}\left(  t\right)  \right\vert \leq2\varepsilon\left\Vert
\omega\right\Vert _{\infty}\exp\left(  2^{-1}t^{2}\sigma_{\left(  m\right)
}\right)  \label{eq16}%
\end{equation}
with probability at least $1-\varepsilon^{-2}p_{m}\left(  t\right)  $. On
other hand, applying H\"{o}lder's inequality to the RHS of (\ref{eqx1}) and
noticing (\ref{eq16}), we have%
\begin{equation}
\mathbb{V}\left[  \hat{\varphi}_{m}\left(  t,\mathbf{z}\right)  \right]
=\mathbb{E}\left[  \left\vert e_{m}\left(  t\right)  \right\vert ^{2}\right]
\leq4\left\Vert \omega\right\Vert _{\infty}^{2}\exp\left(  t^{2}%
\sigma_{\left(  m\right)  }\right)  p_{m}\left(  t\right)  . \label{eqx2}%
\end{equation}
This proves the claims on the oscillations of $\left\vert e_{m}\left(
t\right)  \right\vert $.

Now we show the second claim. Let $t_{\gamma,m}=\sqrt{2\sigma_{\left(
m\right)  }^{-1}\gamma\ln m}$ and $A_{m}\left(  \gamma\right)  =\left(
0,t_{\gamma,m}\right]  $ and $B_{m}\left(  \gamma\right)  =\left[
0,t_{\gamma,m}\right]  $. Since $m^{-2}\left\Vert \mathbf{\Sigma}\right\Vert
_{1}\leq Cm^{-\delta}$ for some constants $C,\delta>0$ for all sufficiently
large $m$ and $\inf_{m\geq1}\sigma_{\left(  m\right)  }>0$, we see%
\[
\sup\limits_{t\in B_{m}\left(  \gamma\right)  }p_{m}\left(  t\right)
=m^{-1}+\sup\limits_{t\in B_{m}\left(  \gamma\right)  }t^{2}m^{-2}\left\Vert
\mathbf{\Sigma}\right\Vert _{1}\leq C^{\prime}m^{-\delta^{\prime}}%
\]
for any $0<\delta^{\prime}<\min\left\{  \delta,1\right\}  $ and some constant
$C^{\prime}>0$. Pick two constants $\eta,\gamma$ such that $0<\gamma<\eta$ and
$2\eta<\delta^{\prime}$, and set $\varepsilon=m^{-\eta}$. Then (\ref{eq16})
implies%
\begin{equation}
\sup_{t\in A_{m}\left(  \gamma\right)  }\left\vert e_{m}\left(  t\right)
\right\vert \leq2\varepsilon\left\Vert \omega\right\Vert _{\infty}\sup_{t\in
A_{m}\left(  \gamma\right)  }\exp\left(  2^{-1}t^{2}\sigma_{\left(  m\right)
}\right)  \leq2\left\Vert \omega\right\Vert _{\infty}m^{\gamma-\eta}.
\label{eq15}%
\end{equation}
This completes the proof.

\subsection{Proof of \autoref{ThmConsistPlugEst}}

The proof is similar to that for Theorem 5 in \cite{Jin:2008}. Recall
$I_{1,m}=\left\{  1\leq j\leq m:|\mu_{j}|>0\right\}  $ and the positive
constants specified in \autoref{LmBoundDiffPhaseFuncWTumix}, i.e., $\eta$,
$\gamma$ and $\delta^{\prime}$ satisfy $0<\gamma<\eta$ and $2\eta
<\delta^{\prime}<\min\left\{  \delta,1\right\}  $. Let $t_{\gamma,m}=$
$\sqrt{2\sigma_{\left(  m\right)  }^{-1}\gamma\ln m}$. Using the decomposition
$\hat{\varphi}_{m}\left(  t_{\gamma,m},\mathbf{z}\right)  -\pi_{1,m}%
=r_{1}+r_{2}$, where $r_{1}=\hat{\varphi}_{m}\left(  t_{\gamma,m}%
,\mathbf{z}\right)  -\varphi_{m}\left(  t_{\gamma,m},\boldsymbol{\mu}\right)
$ and $r_{2}=\varphi_{m}\left(  t_{\gamma,m},\boldsymbol{\mu}\right)
-\pi_{1,m}$, we proceed as follows.

First, we deal with $r_{1}$. By \autoref{LmBoundDiffPhaseFuncWTumix}, we have
$\left\vert \hat{\varphi}_{m}\left(  t_{\gamma,m},\mathbf{z}\right)
-\varphi_{m}\left(  t_{\gamma,m},\boldsymbol{\mu}\right)  \right\vert
\leq2\left\Vert \omega\right\Vert _{\infty}m^{-\left(  \eta-\gamma\right)  }$
with probability at least $1-Cm^{-\left(  \delta^{\prime}-2\eta\right)  }$.
However, $\lim_{m\rightarrow\infty}\pi_{1,m}m^{\eta-\gamma}=\infty$ by
assumption. So,%
\[
\left\vert \frac{r_{1}}{\pi_{1,m}}\right\vert =\left\vert \frac{\hat{\varphi
}_{m}\left(  t_{\gamma,m},\mathbf{z}\right)  -\varphi_{m}\left(  t_{\gamma
,m},\boldsymbol{\mu}\right)  }{\pi_{1,m}}\right\vert \leq\frac{2\left\Vert
\omega\right\Vert _{\infty}m^{-\left(  \eta-\gamma\right)  }}{\pi_{1,m}%
}\rightarrow0
\]
with probability tending to $1$ as $m\rightarrow\infty$.

Now we deal with $r_{2}$. Recall the cardinality of $I_{1,m}$ as $m_{1}$ and%
\[
\psi\left(  t,\mu\right)  =\int_{\left[  -1,1\right]  }\cos\left(  t\mu
s\right)  \omega\left(  s\right)  ds.
\]
Since $\lim_{m\rightarrow\infty}\tilde{u}_{m}\sqrt{2\gamma\ln m}=\infty$ by
assumption, the definitions of $\varphi_{m}$ and $\psi$ imply%
\begin{align*}
\left\vert \frac{r_{2}}{\pi_{1,m}}\right\vert  &  =\left\vert \frac
{\varphi_{m}\left(  t_{\gamma,m},\boldsymbol{\mu}\right)  }{\pi_{1,m}%
}-1\right\vert =\frac{1}{m_{1}}\sum_{j\in I_{1,m}}\left\vert \psi\left(
t,\mu_{j}\right)  \right\vert \\
&  \leq\sup_{\left\{  t\mu:t\mu\geq\tilde{u}_{m}\sqrt{2\gamma\ln m}\right\}
}\left\vert \psi\left(  t,\mu\right)  \right\vert \rightarrow0
\end{align*}
as $m\rightarrow\infty$, where we have applied the Riemann-Lebesgue lemma to
$\psi\left(  t,\mu\right)  $. Therefore, as \ $m\rightarrow\infty$
\[
\mathbb{P}\left(  \left\vert \pi_{1,m}^{-1}\hat{\varphi}_{m}\left(
t_{\gamma,m},\mathbf{z}\right)  -1\right\vert \rightarrow0\right)
\rightarrow1.
\]
This completes the proof.

\subsection{Two auxiliary lemmas}

The following two lemmas, which were proved by \cite{Chen:2019a}, are needed
for the proofs that are to be given in later subsections. Specifically, the first is
on the speed of convergence of Dirichlet integral, stated as

\begin{lemma}
\label{lm:Dirichlet}$\left\vert \int_{0}^{t}x^{-1}\sin xdx-2^{-1}%
\pi\right\vert \leq2\pi t^{-1}$ for $t\geq2$.
\end{lemma}

\noindent and the second is on the speed of convergence of Riemann-Lebesgue
lemma, stated as

\begin{lemma}
\label{lm:OracleSpeed}Let $-\infty<a_{1}<b_{1}<\infty$. If $f:\left[
a_{1},b_{1}\right]  \rightarrow\mathbb{R}$ is of bounded variation, then%
\[
\left\vert \int_{\left[  a_{1},b_{1}\right]  }f\left(  s\right)  \cos\left(
ts\right)  ds\right\vert \leq2\left(  b_{1}-a_{1}\right)  \left(  \left\Vert
f\right\Vert _{\mathrm{TV}}+\left\Vert f\right\Vert _{\infty}\right)
\left\vert t\right\vert ^{-1}\text{ for }t\neq0,
\]
where $\left\Vert f\right\Vert _{\mathrm{TV}}$ is the total variation and
$\left\Vert f\right\Vert _{\infty}$ the essential supremum of $f$ on $\left[
a_{1},b_{1}\right]  $.
\end{lemma}

\subsection{Proof of \autoref{ThmIVaDep}}

Let $\hat{S}_{1,m}\left(  w,y\right)  =m^{-1}\sum_{i=1}^{m}\cos\left\{
w\left(  z_{i}-y\right)  \right\}  $, $S_{1,m}\left(  w,y\right)
=\mathbb{E}\left\{  \hat{S}_{1,m}\left(  w,y\right)  \right\}  $ and
$d_{1,m}\left(  w,y\right)  =\hat{S}_{1,m}\left(  w,y\right)  -S_{1,m}\left(
w,y\right)  $ for $y\in\left[  a,b\right]  $ and $w\in\mathbb{R}$. Let
$x_{i}=\cos\left\{  ts\left(  z_{i}-y\right)  \right\}  $ and $c_{ij}%
=\mathsf{cov}\left[  x_{i},x_{j}\right]  $ for $i\neq j$. We need to bound
$\left\vert c_{ij}\right\vert $. Set $\tilde{z}_{i}=z_{i}-y$. Then $\left(
\tilde{z}_{i},\tilde{z}_{j}\right)  ,i\neq j$ is bivariate Normal. Using
almost identical arguments in Part I of the proof of
\autoref{LmBoundDiffPhaseFuncWTumix}, we can obtain $\left\vert c_{ij}%
\right\vert \leq t^{2}s^{2}\left\vert \sigma_{ij}\right\vert $ for $i\neq j$.
So,%
\begin{equation}
\mathbb{V}\left\{  d_{1,m}\left(  ts,y\right)  \right\}  \leq m^{-1}%
+2t^{2}m^{-2}\left\Vert \mathbf{\Sigma}\right\Vert _{1}. \label{eq17x}%
\end{equation}
Set $p_{m}\left(  t\right)  =m^{-1}+2t^{2}m^{-2}\left\Vert \mathbf{\Sigma
}\right\Vert _{1}$. Let $\hat{\varphi}_{1,m}\left(  t,\mathbf{z}\right)
=m^{-1}\sum_{i=1}^{m}K_{1}\left(  t,z_{i}\right)  $, $\varphi_{1,m}\left(
t,\boldsymbol{\mu}\right)  =\mathbb{E}\left\{  \hat{\varphi}_{1,m}\left(
t,\mathbf{z}\right)  \right\}  $ and $e_{1,m}\left(  t\right)  =\hat{\varphi
}_{1,m}\left(  t,\mathbf{z}\right)  -\varphi_{1,m}\left(  t,\boldsymbol{\mu
}\right)  $. We have%
\[
e_{1,m}\left(  t\right)  =\frac{t}{2\pi}\int_{a}^{b}dy\int_{-1}^{-1}%
\exp\left(  2^{-1}t^{2}s^{2}\sigma^{2}\right)  d_{1,m}\left(  ts,y\right)
ds.
\]
Applying H\"{o}lder's inequality and (\ref{eq17x})\ to $e_{1,m}\left(
t\right)  $ gives%
\begin{align}
\mathbb{V}\left\{  e_{1,m}\left(  t\right)  \right\}   &  \leq\frac{t^{2}%
}{4\pi^{2}}\left(  b-a\right)  ^{2}\int_{0}^{1}2\exp\left(  t^{2}s^{2}%
\sigma^{2}\right)  ds\int_{0}^{1}2\mathbb{V}\left\{  d_{1,m}\left(
ts,y\right)  \right\}  ds\nonumber\\
&  \leq p_{m}\left(  t\right)  \frac{t^{2}\left(  b-a\right)  ^{2}}{\pi^{2}%
}\frac{\exp\left(  t^{2}\sigma^{2}\right)  -1}{t^{2}\sigma^{2}}.
\label{eq17b1}%
\end{align}

Let $\hat{\varphi}_{0,m}\left(  t,\mathbf{z};u\right)  =m^{-1}\sum
_{i=1}^{m}K_{1,0}\left(  t,z_{i};u\right)  $ for $u\in\mathbb{R}$ and
$e_{0,m}\left(  t;u\right)  =\hat{\varphi}_{0,m}\left(  t,\mathbf{z}%
;u\right)  -\mathbb{E}\left\{  \hat{\varphi}_{0,m}\left(  t,\mathbf{z}%
;u\right)  \right\}  $. Then%
\[
e_{0,m}\left(  t;u\right)  =\int_{-1}^{1}\exp\left(  2^{-1}t^{2}s^{2}%
\sigma^{2}\right)  d_{1,m}\left(  ts,u\right)  \omega\left(  s\right)  ds.
\]
Using similar arguments in Part I\ and II of the proof of
\autoref{LmBoundDiffPhaseFuncWTumix}, we can obtain%
\begin{equation}
\mathbb{V}\left\{  e_{0,m}\left(  t;u\right)  \right\}  \leq4\left\Vert
\omega\right\Vert _{\infty}^{2}p_{m}\left(  t\right)  \frac{\exp\left(
t^{2}\sigma^{2}\right)  -1}{t^{2}\sigma^{2}}. \label{eq17c1}%
\end{equation}
Recall $e_{m}\left(  t\right)  =\hat{\varphi}_{m}\left(  t,\mathbf{z}\right)
-\varphi_{m}\left(  t,\boldsymbol{\mu}\right)  $. Inserting (\ref{eq17b1}) and
(\ref{eq17c1}) into the inequality%
\[
\mathbb{V}\left\{  e_{m}\left(  t\right)  \right\}  \leq2\mathbb{V}\left\{
e_{1,m}\left(  t\right)  \right\}  +4\mathbb{V}\left\{  2^{-1}e_{0,m}\left(
t;a\right)  \right\}  +4\mathbb{V}\left\{  2^{-1}e_{0,m}\left(  t;b\right)
\right\}
\]
gives%
\begin{equation}
\mathbb{V}\left\{  e_{m}\left(  t\right)  \right\}  \leq2p_{m}\left(
t\right)  \frac{\exp\left(  t^{2}\sigma^{2}\right)  -1}{t^{2}\sigma^{2}%
}\left\{  \frac{t^{2}\left(  b-a\right)  ^{2}}{\pi^{2}}+4\left\Vert
\omega\right\Vert _{\infty}^{2}\right\}  . \label{eq17d1}%
\end{equation}

Now we show the consistency. Let $\varepsilon>0$ be a fixed constant and
$t_{m}=\sqrt{\gamma\sigma^{-2}\ln m}$ for some $\gamma>0$. Then (\ref{eq17d1})
implies
\begin{align}
\Pr\left\{  \frac{\left\vert e_{m}\left(  t\right)  \right\vert }{\pi_{1,m}%
}>\varepsilon\right\}   &  \leq\frac{Cp_{m}\left(  t\right)  \exp\left(
t^{2}\sigma^{2}\right)  }{\varepsilon^{2}\pi_{1,m}^{2}}\nonumber\\
&  \leq\frac{Ct^{2}m^{-2}\left\Vert \mathbf{\Sigma}\right\Vert _{1}\exp\left(
t^{2}\sigma^{2}\right)  }{\varepsilon^{2}\pi_{1,m}^{2}}=\frac{Cm^{-2}%
\left\Vert \mathbf{\Sigma}\right\Vert _{1}m^{\gamma}\ln m}{\varepsilon^{2}%
\pi_{1,m}^{2}}. \label{eq17e1}%
\end{align}
So, the RHS of (\ref{eq17e1}) converges to $0$ as $m\rightarrow\infty$ when
$m^{-2}\left\Vert \mathbf{\Sigma}\right\Vert _{1}m^{\gamma}\ln m=o\left(
\pi_{1,m}^{2}\right)  $. Consider the decomposition%
\[
\hat{\varphi}_{m}\left(  t_{m},\mathbf{z}\right)  =-e_{1,m}\left(  t\right)
-2^{-1}e_{0,m}\left(  t_{m};a\right)  -2^{-1}e_{0,m}\left(  t_{m};b\right)
+\tilde{r}_{m}=-e_{m}\left(  t\right)  +\tilde{r}_{m},
\]
where%
\[
\tilde{r}_{m}=1-\varphi_{1,m}\left(  t_{m},\boldsymbol{\mu}\right)
+2^{-1}\varphi_{0,m}\left(  t_{m},\boldsymbol{\mu};a\right)  +2^{-1}%
\varphi_{0,m}\left(  t_{m},\boldsymbol{\mu};b\right)  .
\]
So, it is left to show $\left\vert \pi_{1,m}^{-1}\tilde{r}_{m}-1\right\vert
\rightarrow0$. By \autoref{lm:OracleSpeed}, we have%
\[
\left\vert \psi_{1,0}\left(  t,\mu;\mu^{\prime}\right)  \right\vert
\leq4\left(  \left\Vert \omega\right\Vert _{\mathrm{TV}}+\left\Vert
\omega\right\Vert _{\infty}\right)  \left[  \frac{1_{\left\{  \mu^{\prime}%
\neq\mu\right\}  }\left(  \mu,\mu^{\prime}\right)  }{\left\vert t\left(
\mu-\mu^{\prime}\right)  \right\vert }+1_{\left\{  \mu^{\prime}=\mu\right\}
}\left(  \mu,\mu^{\prime}\right)  \right]  ,
\]
and%
\begin{equation}
\max_{u\in\left\{  a,b\right\}  }\max_{\left\{  j:\mu_{j}\neq u\right\}
}\left\vert \psi_{1,0}\left(  t_{m},\mu_{j};u\right)  \right\vert \leq\frac
{1}{t_{m}}\max_{u\in\left\{  a,b\right\}  }\frac{4\left(  \left\Vert
\omega\right\Vert _{\mathrm{TV}}+\left\Vert \omega\right\Vert _{\infty
}\right)  }{\min_{\left\{  j:\mu_{j}\neq u\right\}  }\left\vert \mu
_{j}-u\right\vert }=\frac{C}{t_{m}u_{m}}, \label{eq10aa}%
\end{equation}
where $u_{m}=\min_{u\in\left\{  a,b\right\}  }\min_{\left\{  j:\mu_{j}\neq
u\right\}  }\left\vert \mu_{j}-u\right\vert $. This, together with
\autoref{lm:Dirichlet}, implies%
\begin{align}
\left\vert \pi_{1,m}^{-1}\tilde{r}_{m}-1\right\vert  &  \leq\frac{6\pi}%
{t_{m}\pi_{1,m}}+\frac{1}{2\pi_{1,m}m}\sum_{u\in\left\{  a,b\right\}  }%
\sum\nolimits_{\left\{  j:\mu_{j}\neq u\right\}  }\left\vert \psi_{1,0}\left(
t_{m},\mu_{j};u\right)  \right\vert \nonumber\\
&  \leq\frac{6\pi}{t_{m}\pi_{1,m}}+\frac{C}{t_{m}u_{m}\pi_{1,m}}.
\label{eq10cc}%
\end{align}
Therefore, $t_{m}^{-1}\left(  1+u_{m}^{-1}\right)  =o\left(  \pi_{1,m}\right)
$ forces $\left\vert \pi_{1,m}^{-1}\tilde{r}_{m}-1\right\vert \rightarrow0$,
completing the proof.

\subsection{Proof of \autoref{ThmTypeI-VDep}}

Before we present the arguments, we need the following

\begin{lemma}
\label{LmCovTrans}Let $\left(  X,Y\right)  $ be bivariate Normal with mean
vector $\mu_{XY}=\left(  \mu_{X},\mu_{Y}\right)^{\top}$, correlation $\rho_{XY}$
and respectively positive, standard deviations $\sigma_{X}$ and $\sigma_{Y}$.
If $\left\vert \rho_{XY}\right\vert <1$, then there exists some constant
$K_{0}>0$ such that%
\[
\left\vert \mathsf{cov}\left[  \cos\left(  tX\right)  ,\cos\left(  tY\right)
\right]  \right\vert \leq K_{0}^{2}\left\vert \rho_{XY}\right\vert D\left(
\rho_{XY}\right)
\]
and%
\[
\left\vert \mathsf{cov}\left[  X\cos\left(  tX\right)  ,Y\cos\left(
tY\right)  \right]  \right\vert \leq K_{0}^{2}\left\vert \rho_{XY}\right\vert
\left(  2\sigma_{X}+\sqrt{\pi}\left\vert \mu_{X}\right\vert \right)  \left(
2\sigma_{Y}+\sqrt{\pi}\left\vert \mu_{Y}\right\vert \right)  D\left(
\rho_{XY}\right)  ,
\]
where $D\left(  \rho_{XY}\right)  =\sum_{n=1}^{\infty}\left\vert \rho
_{XY}\right\vert ^{n-1}n^{-1/6}$.
\end{lemma}

The proof of \autoref{LmCovTrans} is given in \autoref{LastLemma}.
\autoref{LmCovTrans} reveals that, if each pair $\left(  z_{i},z_{j}\right)
,i\neq j$ is bivariate Normal and the means and variances of $\left\{
z_{i}\right\}  _{i=1}^{m}$ are uniformly bounded, then the $l_{1}$-norms of
the covariance matrices of $\left\{  z_{i}\right\}  _{i=1}^{m}$, $\left\{
\cos\left(  tz_{i}\right)  \right\}  _{i=1}^{m}$ and $\left\{  z_{i}%
\cos\left(  tz_{i}\right)  \right\}  _{i=1}^{m}$ have the same order.

Now we present the arguments. Set $\hat{\varphi}_{2,m}\left(  t,\mathbf{z}%
\right)  =m^{-1}\sum_{i=1}^{m}K_{1}\left(  t,z_{i}\right)  $ and
$e_{2,m}\left(  t\right)  =\hat{\varphi}_{2,m}\left(  t,\mathbf{z}\right)
-\mathbb{E}\left\{  \hat{\varphi}_{2,m}\left(  t,\mathbf{z}\right)  \right\}
$. First, we derive an upper bound for $\mathbb{V}\left\{  e_{2,m}\left(
t\right)  \right\}  $. Define $\tilde{S}_{2,m}\left(  t,y\right)  =m^{-1}%
\sum_{i=1}^{m}\sin\left(  tyz_{i}\right)  $, $S_{2,m}\left(  t,y\right)
=\mathbb{E}\left\{  \tilde{S}_{2,m}\left(  t,y\right)  \right\}  $ and
$d_{2,m}\left(  t,y\right)  =\tilde{S}_{2,m}\left(  t,y\right)  -S_{2,m}%
\left(  t,y\right)  $ for $t\in\mathbb{R}$ and $y\in\left[  0,1\right]  $.
Then the arguments in Part I\ of the proof of
\autoref{LmBoundDiffPhaseFuncWTumix} imply%
\begin{equation}
\mathbb{V}\left\{  d_{2,m}\left(  ts,y\right)  \right\}  \leq m^{-1}%
+2t^{2}m^{-2}\left\Vert \mathbf{\Sigma}\right\Vert _{1} \label{eq6d}%
\end{equation}
for $s\in\left[  -1,1\right]  $. Set $\tilde{S}_{3,m}\left(  t,y\right)
=m^{-1}\sum_{i=1}^{m}z_{i}\cos\left(  tyz_{i}\right)  $, $S_{3,m}\left(
t,y\right)  =\mathbb{E}\left\{  \tilde{S}_{3,m}\left(  t,y\right)  \right\}  $
and $d_{3,m}\left(  t,y\right)  =\tilde{S}_{3,m}\left(  t,y\right)
-S_{3,m}\left(  t,y\right)  $ for $t\in\mathbb{R}$ and $y\in\left[
0,1\right]  $. Then, for $s\in\left[  -1,1\right]  $ and $y\in\left[
0,1\right]  $ \autoref{LmCovTrans} implies
\begin{align}
\mathbb{V}\left\{  d_{3,m}\left(  ts,y\right)  \right\}   &  \leq\frac
{\sigma^{2}}{m}+\frac{1}{m^{2}}\sum_{i=1}^{m}\mu_{i}^{2}+\frac{1}{m^{2}}%
\sum_{1\leq i<j\leq m}\mathsf{cov}\left[  z_{i}\cos\left(  tsyz_{i}\right)
,z_{i}\cos\left(  tsyz_{j}\right)  \right] \nonumber\\
&  \leq\frac{\sigma^{2}+\mu_{\ast}}{m}+\frac{4K_{0}^{2}\sigma^{2}}%
{m^{2}\left(  \sigma^{2}-c_{\ast}\right)  }\left(  \sigma+\left\vert \mu
_{i}\right\vert \right)  \left(  \sigma+\left\vert \mu_{j}\right\vert \right)
\sum_{1\leq i<j\leq m}\left\vert \sigma_{ij}\right\vert \nonumber\\
&  \leq m^{-1}\left(  \sigma^{2}+\mu_{\ast}\right)  +\frac{4K_{0}^{2}%
\sigma^{2}\left(  \sigma+\mu_{\ast}\right)  ^{2}}{\left(  \sigma^{2}-c_{\ast
}\right)  }m^{-2}\left\Vert \boldsymbol{\Sigma}\right\Vert _{1}, \label{eq6e}%
\end{align}
where $\mu_{\ast}=\max_{1\leq i\leq m}\left\vert \mu_{i}\right\vert $ and
$c_{\ast}=\sup_{m\geq1}\max_{1\leq i<j\leq m}\left\vert \sigma_{ij}\right\vert
$.

Let $p_{m}\left(  t\right)  =m^{-1}+2t^{2}m^{-2}\left\Vert \mathbf{\Sigma
}\right\Vert _{1}$ and%
\[
p_{2,m}=m^{-1}\left(  \sigma^{2}+\mu_{\ast}\right)  +4K_{0}^{2}\sigma
^{2}\left(  \sigma+\mu_{\ast}\right)  ^{2}\left(  \sigma^{2}-c_{\ast}\right)
^{-1}m^{-2}\left\Vert \boldsymbol{\Sigma}\right\Vert _{1}.
\]
Then for the inequality $\mathbb{V}\left\{  e_{2,m}\left(  t\right)  \right\}
\leq2\tilde{I}_{1,m,0}+2\tilde{I}_{1,m,1}$, where%
\[
\tilde{I}_{1,m,0}=\mathbb{E}\left(  \left\{  \frac{1}{2\pi}\int_{0}^{1}%
dy\int_{-1}^{1}\exp\left(  2^{-1}t^{2}y^{2}s^{2}\sigma^{2}\right)
syt^{2}\sigma^{2}d_{2,m}\left(  ts,y\right)  ds\right\}  ^{2}\right)
\]
and%
\[
\tilde{I}_{1,m,1}=\mathbb{E}\left(  \left\{  \frac{1}{2\pi}\int_{0}^{1}%
dy\int_{-1}^{1}t\exp\left(  2^{-1}t^{2}y^{2}s^{2}\sigma^{2}\right)
d_{3,m}\left(  ts,y\right)  ds\right\}  ^{2}\right)  ,
\]
we have%
\begin{align*}
\tilde{I}_{1,m,0}  &  \leq\frac{1}{2\pi^{2}}\left(  \int_{0}^{1}\exp\left(
t^{2}s^{2}\sigma^{2}\right)  t^{4}\sigma^{4}ds\right)  \left(  \int_{-1}%
^{1}\mathbb{V}\left\{  d_{2,m}\left(  ts,y\right)  \right\}  ds\right) \\
&  \leq\pi^{-2}p_{m}\left(  t\right)  t^{2}\sigma^{2}\left\{  \exp\left(
t^{2}\sigma^{2}\right)  -1\right\}
\end{align*}
based on (\ref{eq6d}), and%
\begin{align*}
\tilde{I}_{1,m,1}  &  \leq\frac{t^{2}}{2\pi^{2}}\left(  \int_{0}^{1}%
\exp\left(  t^{2}s^{2}\sigma^{2}\right)  ds\right)  \left(  \int_{-1}%
^{1}\mathbb{V}\left\{  d_{3,m}\left(  ts,y\right)  \right\}  ds\right) \\
&  \leq p_{2,m}\pi^{-2}\sigma^{-2}\left\{  \exp\left(  t^{2}\sigma^{2}\right)
-1\right\}  ,
\end{align*}
based on (\ref{eq6e}). Therefore,%
\begin{equation}
\mathbb{V}\left\{  e_{2,m}\left(  t\right)  \right\}  \leq\frac{\exp\left(
t^{2}\sigma^{2}\right)  -1}{\pi^{2}}\left\{  p_{m}\left(  t\right)
t^{2}\sigma^{2}+\frac{p_{2,m}}{\sigma^{2}}\right\}  . \label{eq13fy}%
\end{equation}

Secondly, we derive an upper bound for $\mathbb{V}\left\{  \hat{\varphi}%
_{m}\left(  t,\mathbf{z}\right)  \right\}  $. Recall $K\left(  t,x\right)
=2^{-1}-K_{1}\left(  t,x\right)  -2^{-1}K_{1,0}\left(  t,x;0\right)  $,
$\hat{\varphi}_{0,m}\left(  t,\mathbf{z};u\right)  =m^{-1}\sum_{i=1}%
^{m}K_{1,0}\left(  t,z_{i};u\right)  $ for $u\in\mathbb{R}$ and
$e_{0,m}\left(  t;u\right)  =\hat{\varphi}_{0,m}\left(  t,\mathbf{z}%
;u\right)  -\mathbb{E}\left\{  \hat{\varphi}_{0,m}\left(  t,\mathbf{z}%
;u\right)  \right\}  $, the last two of which have been defined in the
proof of \autoref{ThmIVaDep}. Then (\ref{eq17d1}) in the proof of
\autoref{ThmIVaDep} says%
\[
\mathbb{V}\left\{  e_{0,m}\left(  t;u\right)  \right\}  \leq4\left\Vert
\omega\right\Vert _{\infty}^{2}p_{m}\left(  t\right)  \frac{\exp\left(
t^{2}\sigma^{2}\right)  -1}{t^{2}\sigma^{2}}.
\]
This, together with (\ref{eq13fy}), implies%
\begin{align}
\mathbb{V}\left\{  e_{m}\left(  t\right)  \right\}   &  \leq2\mathbb{V}%
\left\{  e_{2,m}\left(  t\right)  \right\}  +2\mathbb{V}\left\{  2^{-1}%
e_{0,m}\left(  t;b\right)  \right\} \nonumber\\
&  \leq2\exp\left(  t^{2}\sigma^{2}\right)  \left\{  \frac{t^{2}\sigma^{2}%
}{\pi^{2}}p_{m}\left(  t\right)  +\frac{p_{2,m}}{\sigma^{2}}+\left\Vert
\omega\right\Vert _{\infty}^{2}\frac{p_{m}\left(  t\right)  }{t^{2}\sigma^{2}%
}\right\}  . \label{eq13gy}%
\end{align}

Finally, we prove the consistency. Let $\varepsilon>0$ be a fixed constant and
$t_{m}=\sqrt{\gamma\sigma^{-2}\ln m}$ for some $\gamma>0$. Then (\ref{eq13gy})
implies
\begin{align}
\Pr\left\{  \frac{\left\vert e_{m}\left(  t\right)  \right\vert }{\pi_{1,m}%
}>\varepsilon\right\}   &  \leq\frac{C}{\varepsilon^{2}}\frac{m^{-2+\gamma
}\left\Vert \mathbf{\Sigma}\right\Vert _{1}}{\pi_{1,m}^{2}}\left(  \ln
^{4}m+\mu_{\ast}^{2}+\mu_{\ast}\right)  +\frac{Cm^{\gamma}\mu_{\ast}%
}{\varepsilon^{2}\pi_{1,m}^{2}m}\nonumber\\
&  \leq\frac{C}{\varepsilon^{2}}\frac{m^{-2+\gamma}\left\Vert \mathbf{\Sigma
}\right\Vert _{1}}{\pi_{1,m}^{2}}\left(  \ln^{4}m+\mu_{\ast}^{2}+\mu_{\ast
}\right)  . \label{eq13h}%
\end{align}
So, the RHS of (\ref{eq13h}) converges to $0$ as $m\rightarrow\infty$ when
$m^{-2+\gamma}\left\Vert \mathbf{\Sigma}\right\Vert _{1}\left(  \ln^{4}%
m+\mu_{\ast}^{2}+\mu_{\ast}\right)  =o\left(  \pi_{1,m}^{2}\right)  $.
Consider the decomposition%
\[
\hat{\varphi}_{m}\left(  t_{m},\mathbf{z}\right)  =e_{2,m}\left(  t\right)
+2^{-1}e_{0,m}\left(  t_{m};0\right)  +\tilde{r}_{2,m}=e_{m}\left(  t\right)
+\tilde{r}_{2,m},
\]
where%
\[
\tilde{r}_{2,m}=2^{-1}+\varphi_{2,m}\left(  t_{m},\boldsymbol{\mu}\right)
+2^{-1}\varphi_{0,m}\left(  t_{m},\boldsymbol{\mu};0\right)  .
\]
Then, it suffices to show $\pi_{1,m}^{-1}\tilde{r}_{m}\rightarrow1$. By the
same reasoning that gives (\ref{eq10aa}) and (\ref{eq10cc}), we see that
$t_{m}^{-1}\left(  1+\tilde{u}_{m}^{-1}\right)  =o\left(  \pi_{1,m}\right)  $
with $\tilde{u}_{m}=\min_{\left\{  j:\mu_{j}\neq0\right\}  }\left\vert \mu
_{j}\right\vert $ forces $\pi_{1,m}^{-1}\tilde{r}_{m}\rightarrow1$, completing
the proof.

\subsection{Proof of \autoref{LmCovTrans}}

\label{LastLemma}

Let $\phi\left(  x\right)  =\left(  2\pi\right)  ^{-1/2}\exp\left(
-x^{2}/2\right)  $ and $f_{\rho}$ be the density of standard bivariate Normal
random vector with correlation $\rho\in\left(  -1,1\right)  $, i.e.,%
\[
f_{\rho}\left(  x,y\right)  =\frac{1}{2\pi\sqrt{1-\rho^{2}}}\exp\left\{
-\frac{x^{2}+y^{2}-2\rho xy}{2\left(  1-\rho^{2}\right)  }\right\}  .
\]
Let $H_{0}\left(  x\right)  \equiv1$ and%
\begin{equation}
H_{n}\left(  x\right)  =\left(  -1\right)  ^{n}\frac{1}{\phi\left(  x\right)
}\frac{d^{n}}{dx^{n}}\phi\left(  x\right)  \label{eq6b}%
\end{equation}
be the $n$th Hermite polynomial for $n\in\mathbb{N}_{+}=\{1,2,\ldots\}$, as defined by
\cite{Feller:1971B}. Then Mehler's expansion in \cite{Mehler:1866} implies%
\begin{equation}
f_{\rho}\left(  x,y\right)  =\left\{  1+\sum_{n=1}^{\infty}\frac{\rho^{n}}%
{n!}H_{n}\left(  x\right)  H_{n}\left(  y\right)  \right\}  \phi\left(
x\right)  \phi\left(  y\right)  . \label{eq:Mehler}%
\end{equation}
By \cite{Watson:1933}, the infinite series on the right hand side of
(\ref{eq:Mehler}) as a trivariate function of $\left(  x,y,\rho\right)  $ is
uniformly convergent on each compact set of $\mathbb{R}\times\mathbb{R}%
\times\left(  -1,1\right)  $. Lemma 1 of \cite{Chen:2014SLLN} asserts that
\textquotedblleft for the Hermite polynomials $H_{n}(\cdot)$ defined by
(\ref{eq6b}), there is some constant $K_{0}>0$ independent of $n$ and $y$ such
that
\begin{equation}
\left\vert e^{-y^{2}/2}H_{n}\left(  y\right)  \right\vert \leq K_{0}\sqrt
{n!}n^{-1/12}e^{-y^{2}/4}\text{ \ for any\ }y\in\mathbb{R}."
\label{eqBoundHermite}%
\end{equation}
The constant $K_{0}$ in (\ref{eqBoundHermite}) can be determined using Theorem
1 of \cite{Krasikov:2004} (which refines the upper bound in
(\ref{eqBoundHermite}) for all $n\geq6$) and by computing the upper bounds
explicitly for $e^{-y^{2}/2}\left\vert H_{n}\left(  y\right)  \right\vert $
for $1\leq n\leq5$.

Let $X_{1}=\sigma_{X}^{-1}\left(  X-\mu_{X}\right)  $ and $Y_{1}=\sigma
_{Y}^{-1}\left(  Y-\mu_{Y}\right)  $. Then $\rho_{XY}$ is the correlation
among $X_{1}$ and $Y_{1}$. Set $C_{1}=\mathsf{cov}\left[  \cos\left(
tX\right)  ,\cos\left(  tY\right)  \right]  $. Then%
\[
C_{1}=\mathsf{cov}\left[  \cos\left\{  t\left(  \sigma_{X}X_{1}+\mu
_{X}\right)  \right\}  ,\cos\left\{  t\left(  \sigma_{Y}Y_{1}+\mu_{Y}\right)
\right\}  \right]  ,
\]
and by (\ref{eq:Mehler}) and (\ref{eqBoundHermite}) we have
\begin{align*}
\left\vert C_{1}\right\vert  &  \leq\sum_{n=1}^{\infty}\frac{\left\vert
\rho_{XY}\right\vert ^{n}}{n!}\int\left\vert H_{n}\left(  x\right)
H_{n}\left(  y\right)  \phi\left(  x\right)  \phi\left(  y\right)  \right\vert
dxdy\\
&  \leq \frac{ K_{0}^{2}}{2\pi}\sum_{n=1}^{\infty}\frac{\left\vert
\rho_{XY}\right\vert ^{n}}{n^{1/6}}\leq \frac{ K_{0}^{2}}{2\pi}\left\vert \rho
_{XY}\right\vert \sum_{n=1}^{\infty}\frac{\left\vert \rho_{XY}\right\vert
^{n-1}}{n^{1/6}}.
\end{align*}
Consider $C_{2}=\mathsf{cov}\left[  X\cos\left(  tX\right)  ,Y\cos\left(
tY\right)  \right]  $. Then%
\[
C_{2}=\mathsf{cov}\left[  \left(  \sigma_{X}X_{1}+\mu_{X}\right)  \cos\left\{
t\left(  \sigma_{X}X_{1}+\mu_{X}\right)  \right\}  ,\left(  \sigma_{Y}%
Y_{1}+\mu_{Y}\right)  \cos\left\{  t\left(  \sigma_{Y}Y_{1}+\mu_{Y}\right)
\right\}  \right]  ,
\]
and (\ref{eq:Mehler}) and (\ref{eqBoundHermite}) implies%
\begin{align*}
\left\vert C_{2}\right\vert  &  \leq{K_{0}^{2}} \sum_{n=1}^{\infty}%
\frac{\left\vert \rho_{XY}\right\vert ^{n}}{n!}\int\left\vert \sigma_{X}%
x+\mu_{X}\right\vert \left\vert \sigma_{Y}y+\mu_{Y}\right\vert
\left\vert H_{n}\left(  x\right)  H_{n}\left(  y\right)  \phi\left(  x\right)
\phi\left(  y\right)  \right\vert dxdy\\
&  \leq\frac{K_{0}^{2}}{2\pi}\left(  4\sigma_{X}+2\sqrt{\pi}\left\vert \mu
_{X}\right\vert \right)  \left(  4\sigma_{Y}+2\sqrt{\pi}\left\vert \mu
_{Y}\right\vert \right)  \sum_{n=1}^{\infty}\frac{\left\vert \rho
_{XY}\right\vert ^{n}}{n^{1/6}}\\
&  \leq K_{0}^{2}\left\vert \rho_{XY}\right\vert \left(  2\sigma_{X}+\sqrt
{\pi}\left\vert \mu_{X}\right\vert \right)  \left(  2\sigma_{Y}+\sqrt{\pi
}\left\vert \mu_{Y}\right\vert \right)  \sum_{n=1}^{\infty}\frac{\left\vert
\rho_{XY}\right\vert ^{n-1}}{n^{1/6}},
\end{align*}
where the quantities in the parathetheses in the second inequality are obtained by integrating the upper bound in (\ref{eqBoundHermite}) without $K_0$ and $n$.
This completes the proof.

\section*{Acknowledgements}

I would like to thank Philip T. Reiss and Armin Schwartzman for providing the
brain imaging data set and $\mathsf{R}$ codes used for the analysis in
\cite{Schwartzman:2015} (which was used in an earlier version of the
manuscript), Jiashun Jin for providing the technical report \cite{Jin:2006b}
and warm encouragements, Jianqing Fan for helpful pointers, and Iosif Pinelis
for providing an upper bound on the series $D$ that appears in
\autoref{LmCovTrans}.

\bibliographystyle{dcu}

\end{document}